**Title:** Effects of Non-Compulsory and Mandatory COVID-19 Interventions on Travel Distance and Time Away from Home: The Case of Norway in 2021

**Authors:** Meghana Kamineni (1), Kenth Engø-Monsen (2), Jørgen Eriksson Midtbø (3), Frode Forland (4), Birgitte Freiesleben de Blasio (1 and 3), Arnoldo Frigessi (1), Solveig Engebretsen (5) ((1) Oslo Centre for Biostatistics and Epidemiology. University of Oslo and Oslo University Hospital, (2) Telenor Research, Oslo, Norway, (3) Department of Method Development and Analytics. Norwegian Institute of Public Health, Oslo, Norway, (4) Division of Infection Control, Norwegian Institute of Public Health, Oslo, Norway, (5) Norwegian Computing Center, Oslo, Norway)

**Abstract:**

Background:

Due to the societal, economic, and health costs of COVID-19 non-pharmaceutical interventions (NPIs), it is important to assess their effects. Human mobility serves as a surrogate for human contacts and compliance to NPIs. In Nordic countries, NPIs have mostly been advised and sometimes made mandatory. It is unclear if making NPIs mandatory further reduced mobility.

Aim:

We investigated the effect of non-compulsory and follow-up mandatory measures in major cities and rural regions on human mobility in Norway. We identified NPI categories that most affected mobility.

Methods:

We used mobile phone mobility data from the largest Norwegian operator. We analysed non-compulsory and mandatory measures with before-after and synthetic difference-in-differences approaches. By regression, we investigated the impact of different NPIs on mobility.

Results:

Nationally and in less populated regions, follow-up mandatory measures further decreased time, but not distance travelled. In urban areas, however, follow-up mandates also decreased distance, and the effect exceeded that of initial non-compulsory measures. Stricter metre rules, gyms closing and reopening, restrictions on guests in homes, and face mask recommendations most impacted distance travelled. Time travelled was most affected by gyms closing and restaurants and shops reopening.

Conclusion:

Overall, non-compulsory measures appeared to decrease distance travelled from home, while mandates further decreased this metric in urban areas. Time travelled is reduced more by mandates than by non-compulsory measures for all regions and interventions. Stricter distancing and restricted number of guests were associated with decreases in mobility.

**Comments:**
36 pages, 5 tables, supplement starts on page 20

**Category:** stat.AP

# Introduction

The effects of non-pharmaceutical interventions (NPIs) on COVID-19 transmission are still unclear [1], and assessing their impact is complicated by delays between transmission and measurable changes in COVID-19 cases and hospitalisation, fluctuating adherence, and self-regulatory behaviour. We study the effects of NPIs on human mobility, which serves as a proxy for compliance and contacts beyond households [2].

It is difficult to analyse how NPIs impact infections in observational studies, due to seasonality, new strains, self-regulating behaviour, and multiple interventions with complex interrelations being implemented and/or lifted simultaneously. Moreover, interventions are often lifted during periods with little transmission and few cases, so finding significant differences requires strong effects. Few randomised clinical trials for COVID-19 interventions have been implemented, leading to a lack of evidence [3]. Some studies using observational data have compared similar groups. For instance, one study compared teachers to other working adults to assess the effect of reopening schools in Norway [4].

Other studies have investigated how COVID-19 NPIs affect mobility. By analysing 56 countries, lockdowns and states of emergency were found to reduce mobility [5]. Another study concluded that early curfews backfire by leading to increased mobility during hours outside of curfew restrictions [6]. Other studies have analysed the effects of social distancing interventions on mobility in socioeconomic groups in the US [7] and on geographical regions in the UK [8]. One study compared mobility changes in Nordic countries during the early pandemic and qualitatively connected these differences to interventions [9].

While many countries have implemented mandatory NPIs to curb COVID-19 transmission, Norway, alongside the other Nordic countries, has often employed non-compulsory advice to reduce social contacts and sometimes later made these interventions obligatory [10]. Non-compulsory measures are less invasive and costly than stricter alternatives and have been recommended in other infectious disease pandemics, including influenza [11-13]. However, there is limited knowledge about the effects of COVID-19 non-compulsory NPIs. One study found mobility reductions in Tokyo due to non-compulsory COVID-19 measures [14]. Since non-compulsory measures were often later made mandatory in Norway, we can compare mobility reductions. If follow-up mandates of recommendations do not provide further reductions, we can conclude that mandates did not provide significant additional compliance. We also analyse the effects of specific NPIs, including interventions unique to Norway or less commonly used elsewhere, like limitations of guests in homes, individuals attending events, and serving of alcohol.

One of our main contributions is analysing the effects of NPIs with a novel combination of mobility metrics. Many studies have used data that captures overall movements of individuals relative to a baseline, such as in Google data [15]. We propose alternate metrics, radius of gyration and time away from home, which provide different insights into human behaviour compared to origin-destination mobility data. Radius of gyration describes the distance travelled from home and is useful to study local interventions due to its sensitivity to local effects [16]. It has been used to predict COVID-19 deaths and analyse travel restrictions in Thailand, Tokyo, Austria and Italy [14, 16-19]. A US study analysing stay-at-home orders utilised data about time spent at home, finding that urban populations stayed at home more than rural populations [20]. To our knowledge, our study is the first to analyse the effects of COVID-19 interventions on both distance and time away from home, which should be analysed together as they offer different insights into human behaviour and may not follow similar trends.

# Methods

## Mobility Data and Preprocessing

We utilised mobile phone data from the largest Norwegian operator, Telenor, from 24 January 2021 to 9 January 2022. Each phone is connected to a mobile phone tower, typically the closest location. By tracking towers connected to phones, we can follow movements of devices and their users. We have three mobility metrics per day per individual, aggregated into distributions for each of the 356 municipalities in Norway. The three metrics are the radius of gyration time weighted (meanDistAway), time spent away from home (timeAway), and maximum distance travelled away from home (maxDistAway). meanDistAway measures distance travelled from home, and derivations of all metrics are in the Supplement, section A. For further anonymization, data is unavailable on selected days that occur approximately every three weeks.

We determined daily mean meanDistAway, timeAway, maxDistAway nationally and for each municipality and county when data was available. We utilised these metrics to compare national non-compulsory and follow-up mandatory NPIs and identify effects of intervention categories. To analyse regional NPIs, we accounted for weekday effects by computing relative changes of mean mobility metrics for each day of the week compared to means of the same metric during the same weekday in a reference period. Details are in the Supplement, section B. This normalisation enabled comparison of mobility on consecutive days, which is necessary for synthetic difference-in-differences. Accounting for weekday effects is unnecessary when analysing national NPIs, as we compared week-long periods before and after NPIs were introduced.

## Before-After Analysis

We identified national interventions that comprise non-compulsory measures that were later made mandatory. We compared the following scores for an intervention on day t on national, county, and municipality levels:

$$WeekBefore_t = \frac{1}{7}\sum_{i=t-7}^{t-1} Metric_i \quad (1)$$

$$WeekAfter_t = \frac{1}{7}\sum_{i=t}^{t+6} Metric_i \quad (2)$$

where $Metric_i$ is one of the three mobility metrics defined previously on day $i$. We excluded missing metrics from score calculations. When metrics for one day were missing, we computed the scores using 13 preceding and following days instead of 14 days. In some cases, if time frames for $WeekBefore_t$ or $WeekAfter_t$ overlap with other NPIs, $WeekBefore_t$ or $WeekAfter_t$ is shortened to avoid overlap with other interventions. The sum of the metrics over the time frame is then divided by the number of days in the modified time frame, instead of 7. The weekly mobility change ($WeekMobilityChange_t$) was calculated as a relative percentage change of the mobility.

$$WeekMobilityChange_t = \frac{WeekAfter_t - WeekBefore_t}{WeekBefore_t} * 100$$

As a control analysis, we calculate $WeekMobilityChange_t$ for a date t before the intervention, to understand previous regional mobility trends.

## Synthetic Difference-in-Differences

We identified regional interventions that comprise non-compulsory measures that were later made mandatory, and we analysed mobility changes using the synthetic Difference-in-Differences (SDID) method [21]. SDID compares trends in control locations and time periods to trends in a region with a new intervention to assess the NPI's effect on mobility. We checked that no other major interventions had been added in the control regions during the weeks before and after the NPI.

## Linear Regression

We implemented three regression models to study the associations between $WeekMobilityChange_t$ and the implementation of multiple intervention categories and relative temperature change. We trained the models with data from non-holiday weeks during 2021 for each region of interest, corresponding to when either national or local interventions were introduced, or time points when no new interventions were introduced in the region. The regions analysed are $R$ = {Norway, Oslo, Trondheim, Stavanger, Tromsø, Bergen}, where the latter five are major Norwegian cities.

Each model included 52 time points when interventions were added and 37 mid-points of two week periods with no new interventions between 24 January and 31 December 2021. Supplement, section F includes a mapping of interventions to implementation dates and the algorithm to identify non-intervention time points.

We utilised 22 covariates. All models included a temperature covariate $P_{r,t}$ (TempWeeklyChange), defined as the relative change in temperature from the week before an NPI on day $t$ to the week after in region $r$. All models included 19 intervention category covariates, denoted by $X_{i,r,t}$ which are binary indicators for whether intervention category i was implemented on date t in region r, $r \in R,\ t \in T_r,\ i \in [1,\ n]$ where n is the number of intervention categories. Each model included two interaction covariates, $Int_{A,r,t}$ and $Int_{B,r,t}$, between the weekly temperature change and the two intervention categories A and B that were most significant in an initial regression without interactions. The dependent variable was $Y_{r,t}$, the relative change in mobility from the week before and after an intervention on day t in region r.

Let $\beta = (\beta_0, \beta_1, \dots \beta_{n+1}, \beta_{n+2}, \beta_{n+3})$ be the parameters to estimate and $T_r$ be the intervention and control time points for region $r$. We estimated these coefficients by minimising

$$\widehat{\beta}^* = argmin_{\widehat{\beta} \in R^{n+3}} \{ \sum_{r \in R} \sum_{t \in T_r} ((\beta_0 + \sum_{i=1}^{n} \beta_i * X_{i,r,t} + \beta_{n+1} * Int_{A,r,t} + \beta_{n+2} * I_{B,r,t} + \beta_{n+3} * P_{r,t}) - Y_{r,t})^2 \}. \quad (1)$$

More details are in the Supplement, section F. We reported statistically significant coefficients for the covariates most associated with $WeekMobilityChange_t$ for meanDistAway, timeAway, and maxDistAway.

## Ethical Statement

This study was performed in line with the principles of the Declaration of Helsinki. Since we use anonymized data, approval from an institutional review board or ethics committee is not required.

# Results

The Supplement, section C includes plots of original and normalised daily mean mobility metrics over 2021.

## National Interventions

We identified three national interventions in December 2021. The first comprises only non-compulsory measures, the second comprises mostly non-compulsory measures, while the last includes mandatory measures. We chose 26 November 2021 as a control time point. The interventions and corresponding time frames for the calculation of WeekBefore and WeekAfter are described below.

1. 26 November 2021: Control time point when no interventions were implemented (WeekBefore = 19 November to 25 November, WeekAfter = 26 November to 2 December)
2. 3 December 2021: Recommendation to work from home and reduce close contacts (WeekBefore = 26 November to 2 December, WeekAfter = 3 December to 8 December)
3. 9 December 2021: Recommendation to work from home more, regulations on events, ban on serving alcohol after midnight, and face mask requirement (WeekBefore = 3 December to 8 December, WeekAfter = 9 December to 14 December)
4. 15 December 2021: Mandatory work from home, alcohol serving ban, required digital teaching at universities, and stricter face mask requirements (WeekBefore = 9 December to 14 December, WeekAfter = 2 January to 8 January)

We used before-after analysis to identify the effects of the interventions nationally and in more and less populated regions. We identified the Oslo, Bergen, Trondheim, Tromsø, and Stavanger municipalities as more populated regions. Less populated areas were the half of the municipalities in each county with the least Telenor users on 24 January 2021 (except Oslo, which has one municipality). These municipalities and corresponding number of users are provided in the Supplement, section E.

Nationally, the initial non-compulsory measures effectively reduced mobility (Table 1). All further measures, including mandates on 15 December, minimally affected meanDistAway and maxDistAway. However, making initial measures mandatory on 15 December strongly decreased timeAway, nationally and in all regions.

Table 1 Effects of NPI interventions on mobility nationally and in urban areas
The percentage of mobility change in each of the three mobility metrics from the week before to the week after different intervention dates is shown for selected regions. The intervention dates include a control time point in November and three interventions in December, with only non-compulsory measures on 3 December, mostly non-compulsory measures on 9 December, and follow-up mandatory measures on 15 December. The regions analysed include all of Norway and major cities in Norway.

| Metric | Region | Intervention Date (t) | | | |
|---|---|---|---|---|---|
| | | 26 November | 3 December | 9 December | 15 December |
| meanDistAway | Norway | 3 | -9 | -3 | -2 |
| | Oslo | 10 | -10 | -7 | -15 |
| | Trondheim | 8 | -4 | -8 | -14 |
| | Bergen | 4 | -7 | -2 | -9 |
| | Stavanger | 6 | -15 | -3 | -3 |
| | Tromsø | 19 | -5 | 1 | -1 |
| timeAway | Norway | 8 | -5 | 4 | -7 |
| | Oslo | 6 | -10 | 3 | -13 |
| | Trondheim | 6 | -3 | 4 | -12 |
| | Bergen | 10 | -5 | 2 | -6 |
| | Stavanger | 7 | -5 | 1 | -11 |
| | Tromsø | 9 | -3 | 4 | -8 |
| maxDistAway | Norway | 1 | -9 | -2 | -1 |
| | Oslo | 11 | -8 | -9 | -12 |
| | Trondheim | 0 | -8 | -1 | -16 |
| | Bergen | 4 | -8 | 0 | -7 |
| | Stavanger | -1 | -8 | -6 | -2 |
| | Tromsø | 5 | -3 | -1 | -1 |

In Bergen and Trondheim, follow-up mandates reduced meanDistAway more than initial recommendations. In Tromsø, a less populated city, follow-up mandates did not reduce meanDistAway, while timeAway declined. The recommendations on 9 December reduced meanDistAway in Tromsø and Trondheim more than initial recommendations on 3 December, while both were similarly effective in Bergen. It is difficult to interpret the results from Stavanger and Oslo, as face masks were made mandatory in both cities and home office became required in Oslo on 3 December.

In less populated areas, non-compulsory measures decreased meanDistAway, which was not reduced by follow-up mandates (Table 2). This trend emerged in all counties, except Trøndelag. MeanDistAway was slightly reduced in Nordland, Rogaland, and Vestland, but the magnitudes were small compared to reductions after initial measures.

Table 2 Effects of NPI interventions in less populated regions
The percentage of mobility change in each of the three mobility metrics from the week before to the week after different intervention dates is shown for selected regions. The intervention dates include a control time point in November and three interventions in December, with only non-compulsory measures on 3 December, mostly non-compulsory measures on 9 December, and mandatory follow-up measures on 15 December. The regions analysed include the least populated half of all municipalities in each of 10 counties in Norway. The Oslo county is excluded from this analysis because it only has one municipality.

| Metric | Region | Intervention Date (t) | | | |
|---|---|---|---|---|---|
| | | 26 November | 3 December | 9 December | 15 December |
| meanDistAway | Vestfold og Telemark | 10 | -12 | -1 | 11 |
| | Troms og Finnmark | -14 | -8 | 1 | 3 |
| | Trøndelag | 4 | -11 | 4 | -6 |
| | Viken | 6 | -10 | -1 | 5 |
| | Vestland | 3 | -7 | -5 | -1 |
| | Nordland | 3 | -14 | 4 | -3 |
| | Rogaland | 0 | -14 | 9 | -4 |
| | Møre og Romsdal | 4 | -8 | -4 | 5 |
| | Agder | 7 | -7 | 5 | 2 |
| | Innlandet | 7 | -7 | -2 | 8 |
| timeAway | Vestfold og Telemark | 8 | -3 | 3 | -6 |
| | Troms og Finnmark | 5 | -2 | 4 | -4 |
| | Trøndelag | 9 | -4 | 5 | -8 |
| | Viken | 8 | -5 | 5 | -6 |
| | | | | | |
| | Vestland | 9 | -4 | 4 | -6 |
| | Nordland | 8 | -3 | 4 | -6 |
| | Rogaland | 9 | -5 | 3 | -6 |
| | Møre og Romsdal | 10 | -3 | 4 | -6 |
| | Agder | 10 | -4 | 5 | -7 |
| | Innlandet | 8 | -3 | 5 | -5 |
| maxDistAway | Vestfold og Telemark | 5 | -11 | 0 | 10 |
| | Troms og Finnmark | -13 | -9 | 0 | 2 |
| | Trøndelag | 3 | -11 | 5 | -4 |
| | Viken | 4 | -8 | 0 | 7 |
| | Vestland | 2 | -6 | -4 | -2 |
| | Nordland | 2 | -14 | 2 | -1 |
| | Rogaland | -3 | -12 | 6 | -2 |
| | Møre og Romsdal | 2 | -5 | -3 | 3 |
| | Agder | 4 | -6 | 3 | 1 |
| | Innlandet | 4 | -7 | -2 | 6 |

Mandates decreased timeAway more than initial non-compulsory measures did in less populated areas. These results align with national mobility trends. In all counties, except Vestland and Møre og Romsdal, initial non-compulsory measures were most impactful, and the mostly non-compulsory interventions on 9 December did not further reduce any metrics.

## Regional Interventions

When analysing two Tromsø interventions, we applied the SDID algorithm for 28 October and 9 November 2021 with the Tromsø municipality as the treatment region and Bodø, Harstad, and Trondheim municipalities as control regions. We chose Bodø, the biggest city in Nordland county, and Harstad, the second most populated municipality in the Troms and Finnmark county as of October 2021 [22]. In addition, we included Trondheim to incorporate a larger city to match the urban nature of Tromsø. The non-compulsory measures and mandate decreased meanDistAway, but the former had the largest effect (Table 3). However, only the mandate reduced timeAway and maxDistAway.

Table 3 Effects of regional NPI interventions on mobility
The causal effects of different regional interventions approximated by SDID on each of the three mobility metrics is shown. The interventions analysed comprise pairs of non-compulsory and follow-up mandatory measures from three cities: Tromsø, Bergen, and Trondheim. In the larger cities of Bergen and Trondheim, mandatory measures continued to reduce mean distance and time away from home. In Tromsø, a smaller city, initial non-compulsory measures effectively reduced the mean distance away from home. Further mandatory measures did not reduce distance travelled much more, but did reduce time away from home.

| City | Date | Intervention | meanDistAway | timeAway | maxDistAway |
|---|---|---|---|---|---|
| Tromsø | 10/28 | Recommendation to work at home, reduce social contacts, and use face masks | -0.06 (-0.12, 0.00) | 0 (-0.03, 0.03) | 0.01 (-0.15, 0.17) |
| | 11/9 | Requirement to work at home and use face masks, many events cancelled | -0.03 (-0.03, -0.03) | -0.03 (-0.04, -0.03) | -0.05 (-0.08, -0.01) |
| Bergen | 8/5 | Recommendation to use face masks and limit people in homes to 10 | -0.01 (-0.21, 0.18) | 0.01 (-0.02, 0.05) | 0.04 (0.03, 0.05) |
| | 8/12 | Requirement to use face masks and limit people in homes to 10 | -0.04 (-0.05, -0.03) | -0.02 (-0.06, 0.02) | -0.05 (-0.10, -0.01) |
| Trondheim | 11/2 | Recommendation to use face masks | 0.03 (-0.01, 0.07) | 0.01 (-0.00, 0.02) | -0.02 (-0.05, 0.01) |
| | 11/24 | Requirement to use face masks | -0.05 (-0.06, -0.05) | -0.03 (-0.05, 0.00) | 0.06 (-0.00, 0.13) |

We applied SDID for 5 August and 8 August 2021 with the Bergen municipality as the treatment region. We chose the four largest cities in Norway, Oslo, Stavanger, and Trondheim municipalities as control regions to match the urban nature of Bergen. We applied SDID for 2 November and 24 November 2021 with the Trondheim municipality as the treatment region and Oslo, Stavanger, and Bergen municipalities as control regions to match the urban nature of Trondheim. Non-compulsory measures were ineffective at reducing mobility in Bergen and Trondheim, and mandates were necessary to minimise mobility further (Table 3). Weights and plots of mobility trends detected by SDID are provided in the Supplement, section E.

## Effects of Intervention Categories

Table 4 shows statistically significant features associated with WeeklyMobilityChange in meanDistAway, timeAway, and maxDistAway.

Table 4: NPI intervention categories with largest effects on meanDistAway
The effect of different intervention categories on the changes in meanDistAway, timeAway, and maxDistAway from the week before to the week after an intervention are shown. Intervention categories that had no significant effects on any mobility metrics include measures related to work from home, schools, serving of alcohol, and face mask requirements. The R-squared values for the models predicting changes in meanDistAway, timeAway, and maxDistAway were 0.419, 0.494, and 0.406 respectively. We computed residuals for each model and found no significant regional variations.

| Intervention Category | meanDistAway | | timeAway | | maxDistAway | |
| --- | --- | --- | --- | --- | --- | --- |
| | Coefficient (95% Confidence Interval) | P-Value | Coefficient (95% Confidence Interval) | P-Value | Coefficient (95% Confidence Interval) | P-Value |
| One or two metre rule was implemented. (StricterMeterRule) | -0.19 (-0.34, -0.04) | 0.015 | NS | NS | -0.14 (-0.29, -0.00) | 0.049 |
| The limit on guests within a private home was decreased. (PHDecr) | -0.13 (-0.26, -0.01) | 0.037 | NS | NS | NS | NS |
| Gyms closed. (GymsClosed) | 0.29 (0.13, 0.46) | 0.001 | -0.08 (-0.15, 0.00) | 0.052 (**) | 0.19 (0.04, 0.35) | 0.014 |
| Gyms reopened. (GymsReopen) | -0.13 (-0.26, -0.01) | 0.040 | NS | NS | -0.13 (-0.25, -0.02) | 0.023 |
| Face masks are recommended. (FMRec) | 0.13 (0.03, 0.24) | 0.016 | NS | NS | 0.09 (-0.01, 0.19) | 0.068 |
| More restrictions are placed on events. (EventNumDecr) | NS | NS | NS | NS | -0.08 (-0.16, 0.00) | 0.063 (**) |
| Restaurants, shops, and businesses reopened. (ResShopReopen) | NS | NS | 0.08 (0.03, 0.12) | 0.004 | NS | NS |

** almost statistically significant; NS = not significant at 5% level

NPIs that significantly decreased meanDistAway were stricter metre rules, gyms reopening, and restrictions on guests in homes. Face mask recommendations increased meanDistAway, as people may feel safer with recommendations in place. Gyms closing increased meanDistAway, but decreased timeAway, as people may train less frequently but train outside. Restaurant and shop reopenings increased timeAway. Additionally, event restrictions decreased maxDistAway, but did not significantly affect meanDistAway. Overall, meanDistAway and maxDistAway were influenced by similar categories. Interventions related to work from home, schools, serving of alcohol, and face mask requirements did not significantly affect any mobility metrics.

# Discussion

Table 5 provides an overview of our results to help guide this discussion.We found that initial national non-compulsory measures in December 2021 in Norway reduced distance and time away from home. Follow-up mandates further decreased time away from home. Interestingly, we observed differences between urban and rural regions. In urban areas, follow-up mandates decreased all mobility metrics, while in less populated areas, they only reduced time away from home.

Table 5 Summary of effects of national and regional non-compulsory (NC) and follow-up mandates on distance and time away from home in more and less densely populated areas.
The downward arrow indicates a decrease in the corresponding metric, while the dash signifies no decrease in the metric. The national non-compulsory measures significantly reduced all mobility metrics in all areas. The regional non-compulsory measures reduced mean distance away from home only in less populated areas, but did not significantly lower mobility in general. Both national and regional mandatory measures effectively reduced time away from home in all regions, but only lessened mean time away from home in more populated areas. The effect of non-compulsory measures and follow-up mandatory measures is not consistent between national and regional interventions. However this metric is more variable than the other two metrics, as shown in the Supplement, section C, and therefore likely less indicative of adherence.

| Mobility Metric | National Interventions | | | | Regional Interventions | | | |
|---|---|---|---|---|---|---|---|---|
| | Major Cities (Bergen, Trondheim, Oslo) | | Less Populated Regions + Overall | | Major Cities (Bergen, Trondheim) | | Less Populated Regions (Tromsø) | |
| | NC | Follow-Up Mandate | NC | Follow-Up Mandate | NC | Follow-Up Mandate | NC | Follow-Up Mandate |
| meanDist Away | ↓ | ↓ | ↓ | — | — | ↓ | ↓ | — |
| timeAway | ↓ | ↓ | ↓ | ↓ | — | ↓ | — | ↓ |
| maxDist Away | ↓ | ↓ | ↓ | — | — | — | — | ↓ |

Analysis of separate regional interventions provided slightly different results from analysis of national measures. Still, the regional results support the conclusion that follow-up mandates reduced distance and time travelled in urban areas, except in rural areas, where only time away from home was shorter.We observed that follow-up mandates in large cities decreased distance travelled more than initial recommendations did for all regional interventions analysed. In contrast to the national interventions, regional non-compulsory measures did not reduce mobility in urban areas, but follow-up mandates were found to be effective. In addition, maximum distance travelled was affected differently by regional interventions, as it was reduced by mandates in Tromsø, but not in larger cities. While this contrasts with the national results, it is more important to focus on mean distance and time travelled, as maximum distance is less indicative of compliance as it is a more variable metric. In summary, non-compulsory national measures limited distance and time away from home, while regional non-compulsory measures only reduced distance in less populated areas. Compulsory follow-up measures reduced time travelled further, especially in more populated areas, where distance travelled was further minimised.

Moreover, we observed higher relative reductions of distance and time travelled in more populated areas. This aligns with findings on mobility reduction from compulsory social distancing policies in the UK and ten Western Pacific countries and areas, which identified greater mobility reductions in high density areas [23-24]. One explanation for this difference is that living in rural areas is associated with considerable everyday mobility. Workplaces, retail facilities and key services are often located far from home. In contrast, people in cities live closer to workplaces and shopping centres and have better curbside pick-up and home delivery options. This may explain the smaller effect on distance metrics for rural versus urban inhabitants.

Follow-up mandates decreased time travelled more than non-compulsory measures did in all cases. One plausible explanation is mandatory home office, which increases time spent at home for those noncompliant with initial recommendations. Since follow-up mandates do not further reduce distance travelled in less populated areas, non-compulsory measures may be more appropriate to generate compliance, as non-compulsory measures are less costly and invasive. Strong public trust in the government may explain the effectiveness of non-compulsory measures in Norway [25-26].

In Tromsø and Trondheim, the second set of recommendations in December 2021 was more impactful than the initial measures in March 2021, while the recommendations had similar effects in Bergen. This could be because Tromsø and Trondheim, in contrast with Bergen, had recently experienced peaks in COVID-19 cases. Knowing that cases were declining, the population may have perceived their personal risk to be lower, so non-compulsory measures may have been less effective [27]. Moreover, Tromsø implemented face mask recommendations before 3 December 2021, which can explain why the 3 December interventions minimally affected mobility. This varying effect of non-compulsory measures indicates that recent COVID-19 case trends should be analysed before introducing such measures.

As this is an observational study, our results should be interpreted with caution. With the before-after analysis for national interventions, potential confounders that affect mobility include temperature and weather. While we analyse mobility trends before interventions are enacted, there are no parallel controls. Yet, because we studied short periods before and after NPIs, the assumption of constant differences in absence of interventions is more likely to hold. For the SDID approach, a case-control analysis, it is difficult to choose appropriate control regions. However, our hypothesis about the varying regional effect of follow-up mandates was supported by the analyses using different interventions and two methodological approaches, increasing confidence in our conclusions. A previous study on assessing compliance to COVID-19 interventions using mobility data identified substantial differences in conclusions drawn from raw and normalised mobility data with respect to a reference period [2]. Our results incorporate both normalised and raw mobility and are therefore more robust.

In our analysis, we identified the effect of isolated intervention categories on mobility, which can be connected to effects on reducing transmission. For instance, another study showed that limiting individuals in gatherings to 10 reduced transmission [1], similar to the intervention categories of restricted guests in homes and individuals in events studied here. We found that these two intervention categories decreased mean and max distance travelled respectively, and hence indicate that the reduction in transmissibility identified in [1] could in part be explained by reduction in mobility after restrictions. For the interventions found to not have significant effects, further investigation is necessary to determine validity, but our results provide some evidence that their effect on mobility in Norway may be minimal. NPIs with non-significant effects on all mobility metrics included measures related to teleworking, schools, serving of alcohol, and face mask requirements. Yet, this does not indicate that these measures were not effective as infection control measures, but that they are not associated with mobility.

Another important finding is that interventions affect distance travelled and time spent away from home differently. Nationally, follow-up mandates impacted time travelled more than distance, and these metrics were influenced by very different interventions: stricter metre rules, restrictions on guests in home, and face mask recommendations influenced distance, while restaurants and shops reopening significantly affected time. Future work should investigate which metric is more relevant to COVID-19 transmission and if interventions should be designed to reduce one metric.

We did not assess how interventions impacted the incidence of COVID-19, due to various factors, including under-reporting of cases and delays between transmission and testing positive for COVID-19 [27-28]. Mobility data serves as an early signal for the effects of NPIs. However, limitations with mobile phone mobility data exist, as it is only a proxy for reduction in contact rate and there may be a selection bias of individuals carrying a mobile phone. However, one study found that this selection bias does not drastically affect mobility estimates [29-30]. We also lack mobility data for one day approximately every three weeks due to data anonymization.

Our results are for Norway, a country where the population has high trust in the government and local authorities [25]. This could be similar in other countries, especially in the Nordic countries. However, future studies are necessary to assess whether the results are transferable to other populations and settings.

# Conclusions

Non-compulsory measures reduced distance travelled in less populated areas, and follow-up mandates in more populated areas further decreased distance travelled. We found that stricter metre rules, restrictions on guests in homes, and restaurants and shops reopening were significantly associated with changes in mobility. These observations have important policy implications on which NPIs to implement, the choice between non-compulsory or mandatory measures, and utilising regional or national interventions. Since follow-up compulsory mandates only have minor effects on distance travelled in less populated areas, less invasive and costly non-compulsory measures may be sufficiently effective for rural areas in the case of Norway.

# Acknowledgements

We acknowledge the work of many colleagues at the Norwegian Institute of Public Health for the collection and preparation of the data. We thank Telenor Norway for providing aggregated mobility data, and in particular Kristian Lindalen Stenerud for help with implementation.

# Authors' Contributions

MK, AF, BFDB, SE, KEM, and JEM were involved in project ideation, design and research discussion. FF discussed the concepts, scope and purpose of the study. JEM designed and implemented the mobility data aggregation algorithms. KEM enabled the data sharing, post-processing of the mobility data, and dataset updates. MK identified the interventions of interest and appropriate methods, implemented the data analyses, and wrote the first draft of the manuscript. AF was involved in supervision, and SE made significant contributions to editing the manuscript draft. All authors edited and reviewed the manuscript.

# Supplementary material: Effects of Non-Compulsory and Mandatory COVID-19 Interventions on Travel Distance and Time Away from Home: The Case of Norway in 2021

## A. Aggregated Mobility Data

The analysis in this work is based on mobile phone data aggregated into three distinct measures: radius of gyration, time away from home, and maximum distance away from home. Each of these metrics is calculated for each individual subscriber each day, and is then aggregated into a probability distribution for each municipality and each day. This aggregation is performed automatically inside the mobile network operator's facilities, immediately after data acquisition. The resulting probability distributions are the only information that is stored. Data is not available approximately one day every three weeks due to anonymization purposes. These days, all pseudonyms that identify individuals are replaced, and since individuals' homes and movements cannot be linked, data is not reported.

### I. Radius of Gyration

The radius of gyration for one individual for a day is defined by the formula

$$R_g = \sqrt{\frac{\sum_{i=1}^{N} m_i (\vec{r}_i - \vec{r}_C)^2}{\sum_{i=1}^{N} m_i}}$$

Here, $\vec{r}_i$ contains the coordinates (using the coordinate system UTM33) of the $i$th visited place of an individual, while $\vec{r}_C$ is the "centre of mass", $\vec{r}_C = \sum_i m_i \vec{r}_i / \sum_i m_i$. For meanDistAway, the temporal duration of stay at each $\vec{r}_i$ was used as the weighting parameter $m_i$.

Individuals are assigned to a municipality based on where they were connected at 4 a.m., i.e. where they presumably spent the night. N is the number of locations visited, and is different for every individual and day.

### II. Time Away From home

The time away from home is defined as the amount of time a subscriber spends connected to a cell tower other than their home tower. Their home tower is taken to be where they were connected at 4 a.m., i.e. where they presumably spent the night.

### III. Max Distance Away From home

The max distance away from home is taken as the maximum Euclidean distance between the home tower (as defined in the previous paragraph) and all other towers to which the subscriber is connected during the day.

## B. Normalisation of Mobility Data

We compute the mean of each metric for each day of the week during a three week reference period from September 2 to 23, 2021 when mobility levels were stable and few interventions were in place. Let dayOfWeek be Monday, Tuesday, … or Sunday. Then we have

$$RefMeanDistAway_{dayOfWeek} = \frac{\sum_{t = Sept\ 2}^{Sept\ 23} meanDistAway_t * 1_{t\ is\ a\ dayOfWeek}}{\sum_{t = Sept\ 2}^{Sept\ 23} 1_{t\ is\ dayOfWeek}}$$

$$RefTimeAway_{dayOfWeek} = \frac{\sum_{t = Sept\ 2}^{Sept\ 23} timeAway_t * 1_{t\ is\ a\ dayOfWeek}}{\sum_{t = Sept\ 2}^{Sept\ 23} 1_{t\ is\ a\ dayOfWeek}}$$

$$RefMaxDistAway_{dayOfWeek} = \frac{\sum_{t = Sept\ 2}^{Sept\ 23} maxDistAway_t * 1_{t\ is\ a\ dayOfWeek}}{\sum_{t = Sept\ 2}^{Sept\ 23} 1_{t\ is\ a\ dayOfWeek}}$$

Then, on each day t and for each metric, we computed the relative change of the mean for each day of the week compared to the mean of the same mobility metric during the same weekday in the reference period. Let dayOfWeek(t) be the day of the week of day t.

$$NormMeanDistAway_t = \frac{meanDistAway_t - RefMeanDistAway_{dayOfWeek(t)}}{RefMeanDistAway_{dayOfWeek(t)}}$$

$$NormTimeAway_t = \frac{timeAway_t - RefTimeAway_{dayOfWeek(t)}}{RefTimeAway_{dayOfWeek(t)}}$$

$$NormMaxDistAway_t = \frac{maxDistAway_t - RefMaxDistAway_{dayOfWeek(t)}}{RefMaxDistAway_{dayOfWeek(t)}}$$

We utilised these relative mobility values in the SDID approach, as metrics for consecutive days must be comparable with each other. This normalisation is not necessary for the before-after analysis, as the time periods of comparison are weekly units, rather than daily units as in the SDID analysis, so weekday variation will not affect the analysis.

## C. Mobility Data

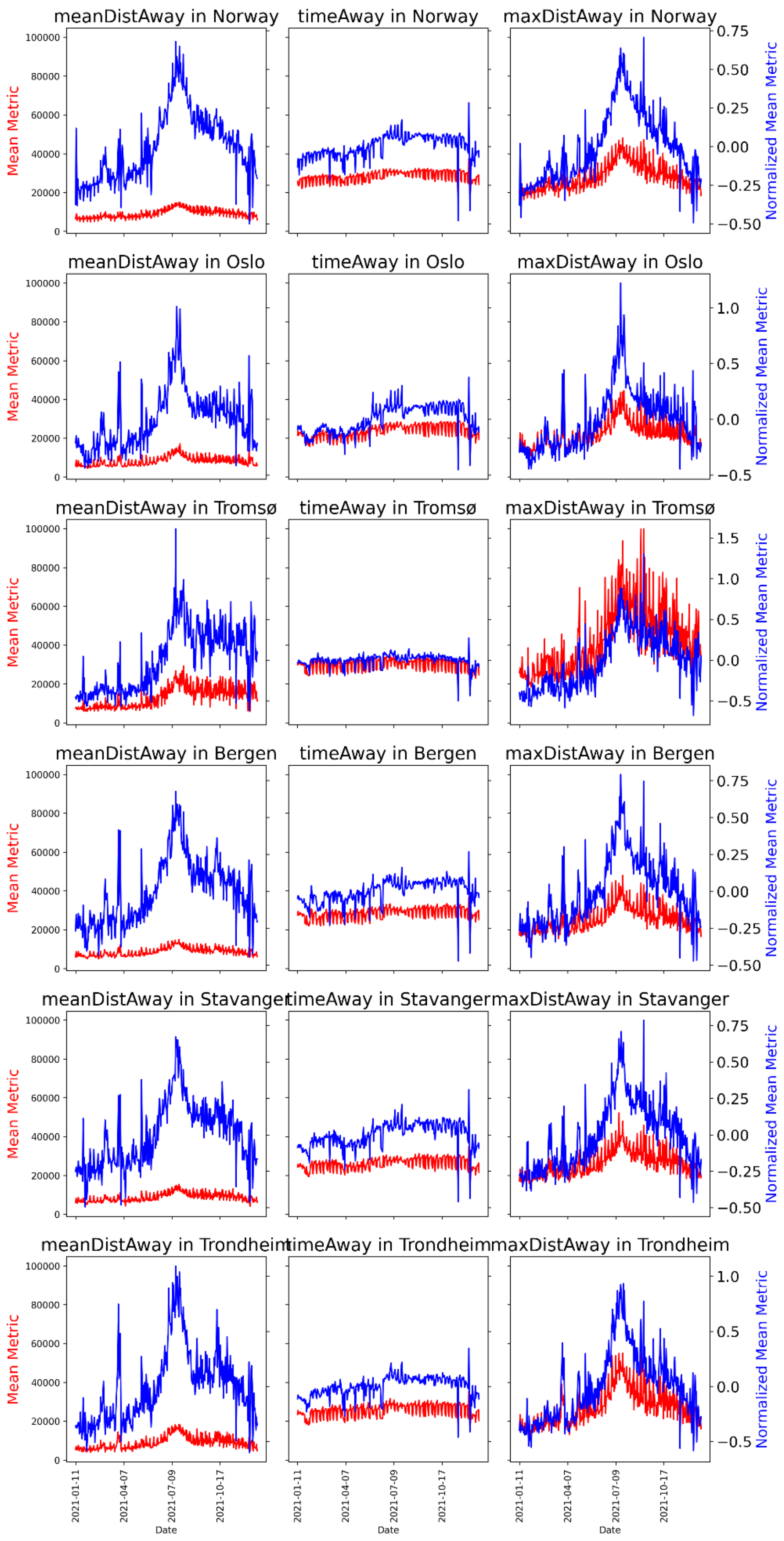

**Figure S1.** Original and normalised mean mobility metrics in Norway and selected cities
The original mean of each mobility metric across the pandemic is plotted in red, while the normalised mean mobility metrics are plotted in blue.

## D. National and Local Interventions

**Table S1.** Non-pharmaceutical interventions in Norway, Oslo, Bergen, Trondheim, Tromsø, and Stavanger from February 1, 2021 to December 31, 2021
This list does not include entry restrictions into the country from abroad or vaccine interventions.

| Geographical Area | Date of Effect | Summary of Action |
|---|---|---|
| Norway | February 3 | More people are allowed in private homes, and event restrictions are eased. [1] |
| | March 25 | A 2-metre rule and face masks are recommended. Alcohol is prohibited, and more restrictions are placed on events. [2] |
| | April 16 | Alcohol, event, school, and private home restrictions are eased in the 1st phase of reopening. [3] |
| | May 27 | Alcohol and event restrictions are eased in the 2nd phase of reopening. [4] |
| | June 20 | Alcohol, event, work-from-home, and private home restrictions are eased in the third phase of reopening, along with the gradual reopening of the travel network. [5] |
| | September 4 | Event restrictions are eased. [6] |
| | September 25 | Alcohol and event restrictions are eased, social distancing recommendations are removed. [7] |
| | December 3 | Work from home and reducing close contacts is recommended. [8] |
| | December 9 | Recommendations to work from home even more are added, along with initial alcohol, event, and face mask restrictions. [9] |
| | December 15 | Work from home becomes mandatory, and alcohol is prohibited. Universities are required to do digital teaching, and the requirement to wear face masks indoors is extended to indoor events, libraries, museums, etc. [10] |
| Oslo | February 3 | Restaurants and shops can reopen, and kindergartens, primary and secondary schools will change from red to yellow risk levels. [11] |
| | March 2 | Restaurants and shops are closed, and event, private home, and school restrictions are added. [12] |
| | April 19 | School restrictions are eased. [13] |
| | May 6 | Restaurants and shops can reopen.[14] |
| | May 26 | Restrictions on alcohol and events are eased, and restaurants and shops can reopen further. [15] |
| | June 16 | 2-metre rule and alcohol, event, and private home limits are eased. [16] |
| | July 5 | Working from home and alcohol restrictions are eased.[17] |
| | November 24 | Face masks are recommended. [18] |
| | December 15 | Red and yellow levels are added for kindergartens and schools in Oslo. [19] |
| Bergen | February 7 | Alcohol, event, school, work from home, and private home restrictions are added, and face masks are required. [20] |
| | February 21 | Alcohol, event, school, work from home, and private home restrictions are eased. [21] |
| | March 27 | Gyms are closed. [22] |
| | March 31 | Gyms opened. [23] |
| | April 19 | Event, school, and private home restrictions are added, and face masks are required. [24] |
| | May 12 | Alcohol restrictions are tightened, but private home restrictions are eased. [25] |
| | May 31 | Event, alcohol, work from home, and private home restrictions are eased. [26] |
| | August 5 | Private home limits up until 10 people and face masks are recommended. [27] |
| | August 12 | Private home limits up until 10 people and face masks are made mandatory. [28] |
| | September 1 | Private home and face mask restrictions are eased. [29] |
| | December 9 | Face masks are required. [30] |
| | December 15 | Schools become digital. [31] |
| Trondheim | February 2 | Alcohol restrictions are eased. [32] |
| | February 10 | Alcohol and work-from-home restrictions are added, and some businesses are closed. [33] |
| | May 19 | Face masks are required. [34] |
| | June 1 | Alcohol, private home, and work from home restrictions are added. [35] |
| | June 22 | Alcohol, private home, work from home, and face mask restrictions are eased. [36] |
| | August 26 | Face masks are required.[37] |
| | September 2 | Events are more restricted, as NTNU recommends smaller student events. [38] |
| | September 21 | Face mask requirement is eased. [39] |
| | November 2 | Face masks are recommended. [40] |
| | November 24 | Face masks are required. [41] |
| Tromsø | March 5 | Face masks are required, gyms are closed, and private home restrictions are added. [42] |
| | March 19 | Face mask and private home restrictions are eased, and gyms re-opened. [42] |
| | July 2 | Face masks are recommended. [43] |
| | July 13 | Face mask recommendation is eased.[44] |
| | Oct 28 | Recommendations to work from home more, reduce social contacts, use face masks, and follow the 1-metre rule are added. [45] |
| | November 9 | Mandatory work from home, obligations to use face masks, and event restrictions are added. [46] |
| | November 30 | Some face mask and event restrictions are eased. [47] |
| Stavanger | April 16 | Event restrictions and mandatory home office are added. [48] |
| | May 6 | Work from home, alcohol, and event restrictions are eased. [49] |
| | July 8 | Event restrictions are eased. [50] |
| | September 1 | Face masks are required in public transport and indoors when not possible to keep a metre distance. [51] |
| | September 15 | Face mask restrictions are eased. [51] |
| | December 2 | Face masks are required. [52] |
| | December 9 | School restrictions are added. [53] |
| | December 15 | School restrictions are added. [54] |

## E. Non-compulsory Measures and Mandates

### I. SDID Weights and Plots of Trends for Regional Interventions

In order to compare trends between the control regions and the region with the intervention (referred to as the treated unit), we used SDID to calculate weights for each control region to create a weighted average of the control regions that is as similar as possible to the treated unit. We present the weights for each control region for each application of SDID to improve interpretability. In addition, we created plots comparing the trend in the synthetic control, or the weighted average of the controls, and in the treated unit.

**Table S2.** SDID Weights for Analysing Tromsø Interventions. The weights assigned to each control region are presented for the applications of SDID to estimate the effect of the two interventions in Tromsø on each of the three mobility metrics.

| Date | Control Region | meanDistAway | timeAway | maxDistAway |
|---|---|---|---|---|
| Oct 28 | Bodø | 0.329 | 0.349 | 0.303 |
| | Harstad | 0.343 | 0.320 | 0.318 |
| | Trondheim | 0.328 | 0.331 | 0.379 |
| Nov 9 | Bodø | 0.332 | 0.277 | 0.348 |
| | Harstad | 0.385 | 0.364 | 0.309 |
| | Trondheim | 0.282 | 0.359 | 0.343 |

**Table S3.** Visualisation of Mobility Trends in Synthetic Control versus Treated Regions for Analysing Tromsø Interventions. The normalised mobility metrics are plotted for a week before and after each intervention for both the synthetic control and treated regions. The synthetic control is a weighted average of the control regions, where the weights are chosen to minimise the difference between the synthetic control and treated region's mobility trends before the intervention.

| | 10/28 - Non-Compulsory Measures | 11/9 - Mandate |
|---|---|---|
| Mean Dist Away | 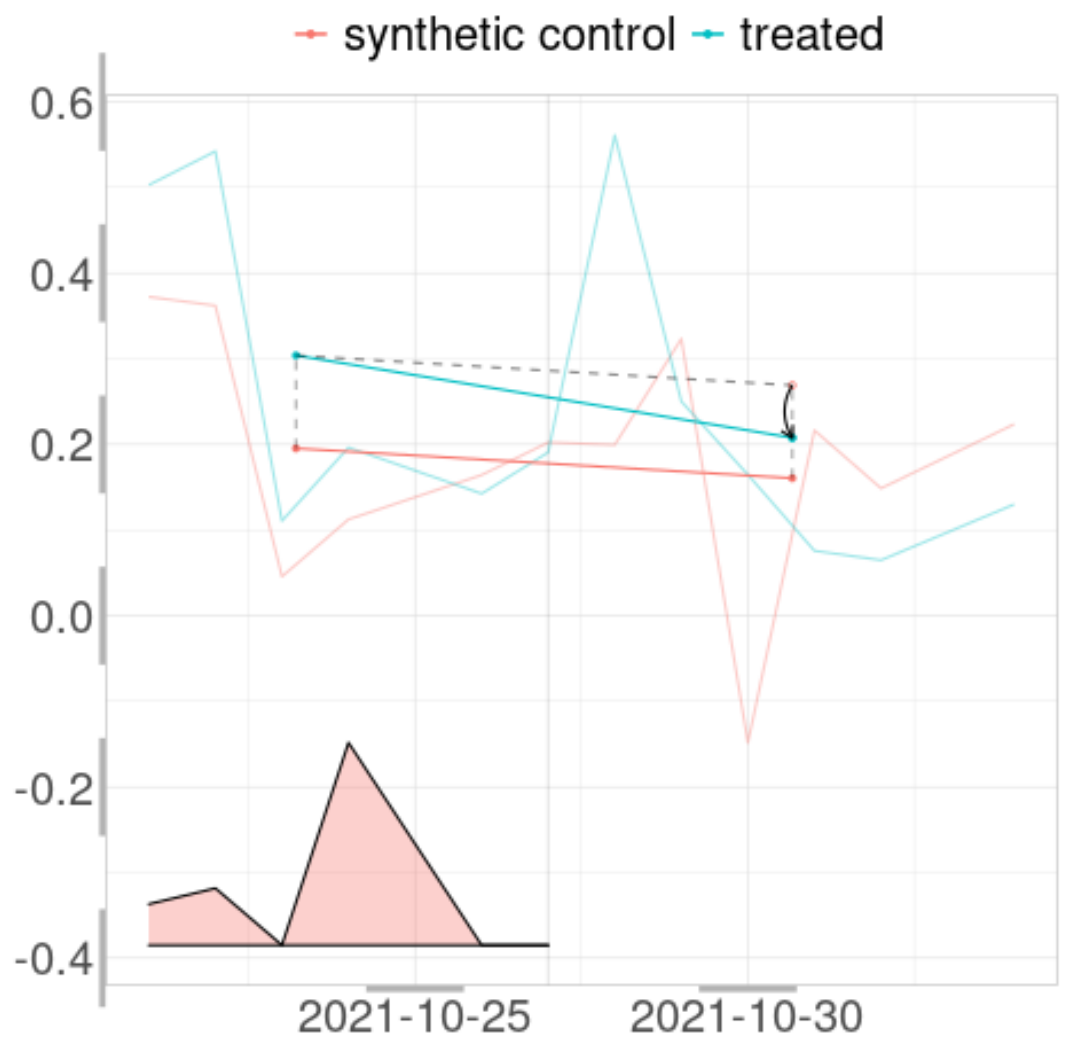  | 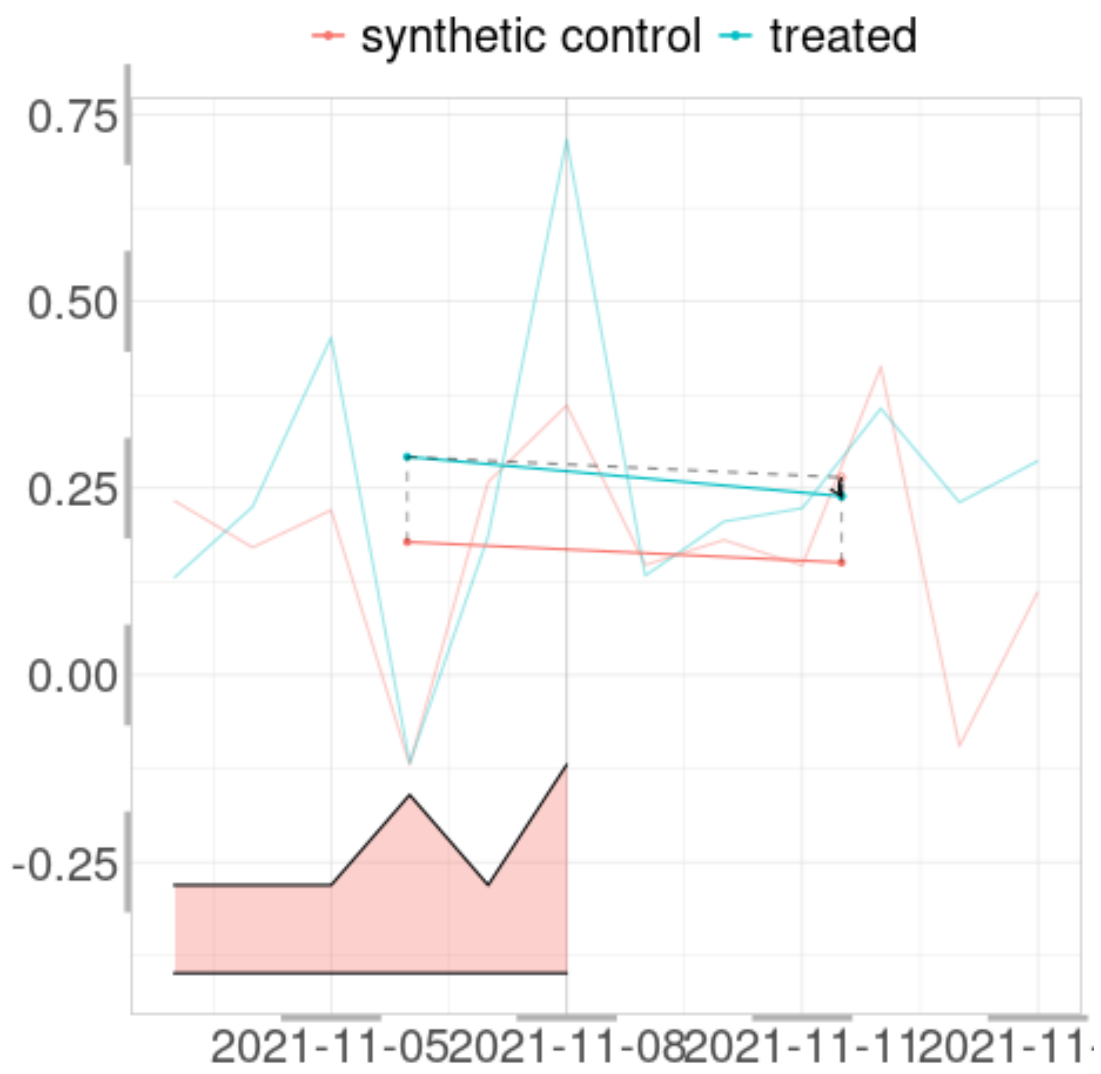  |

<table>
<tr><td>Time Away</td><td>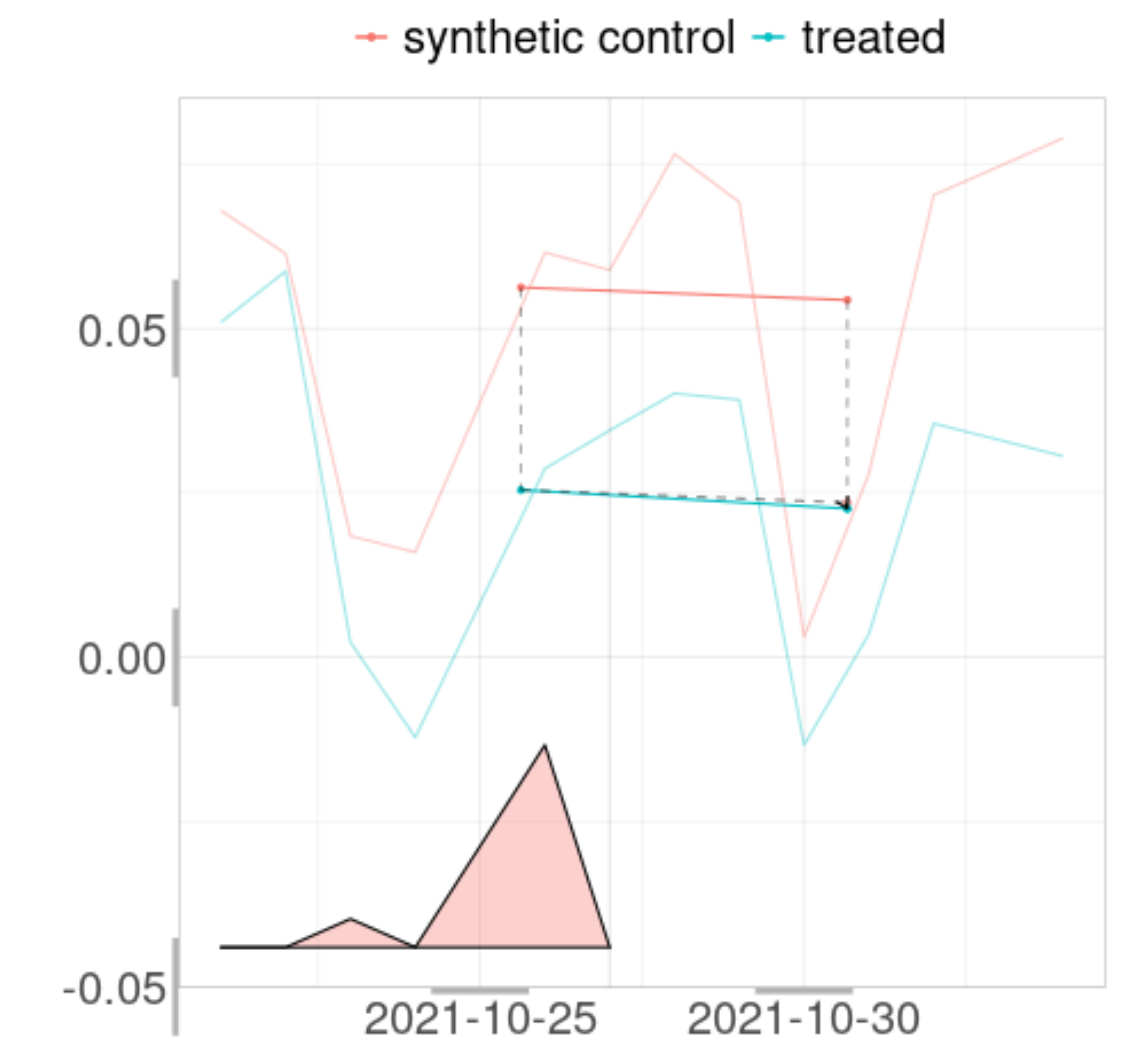
</td><td>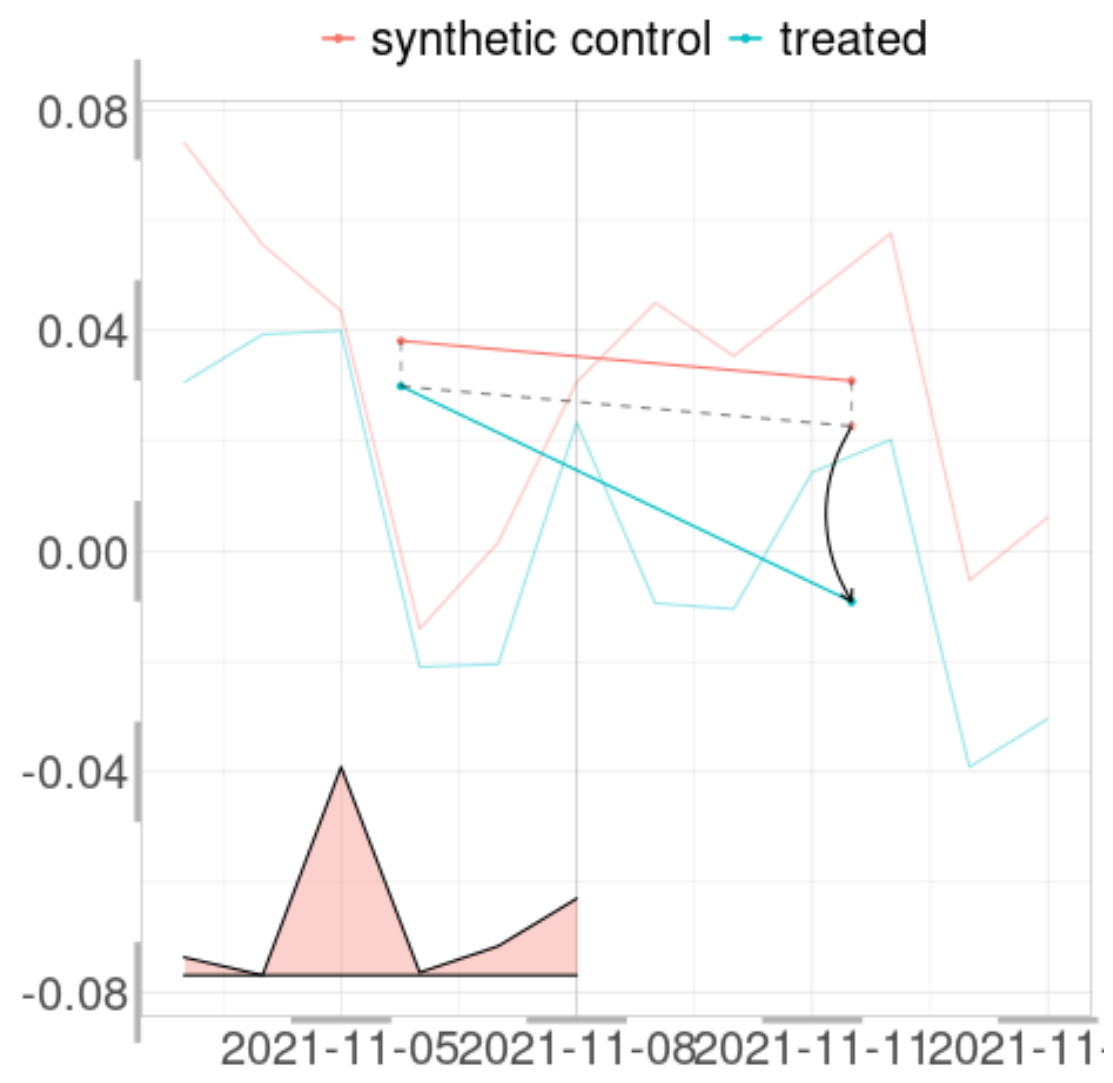
</td></tr>
<tr><td>Max Dist Away</td><td>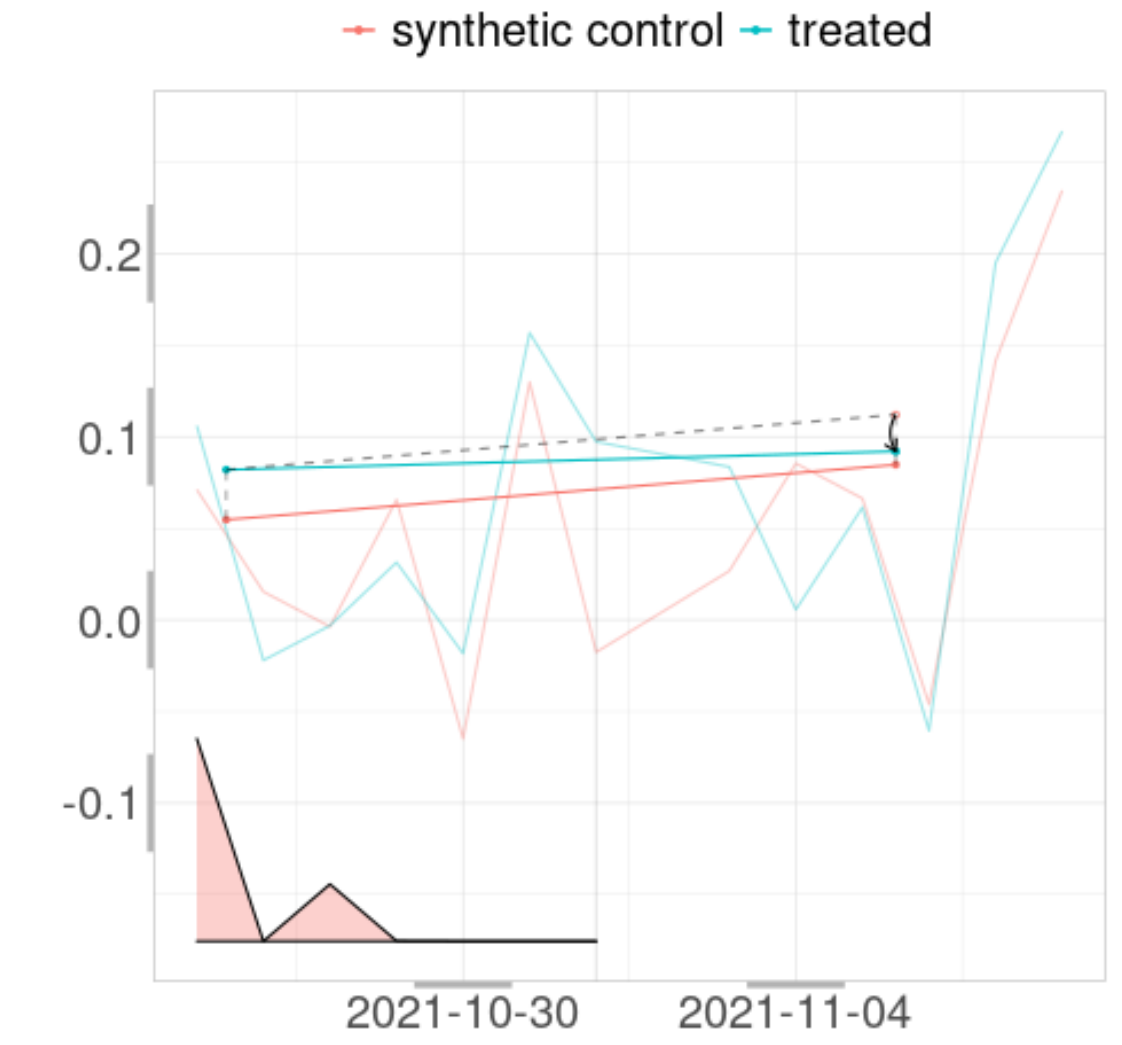
</td><td>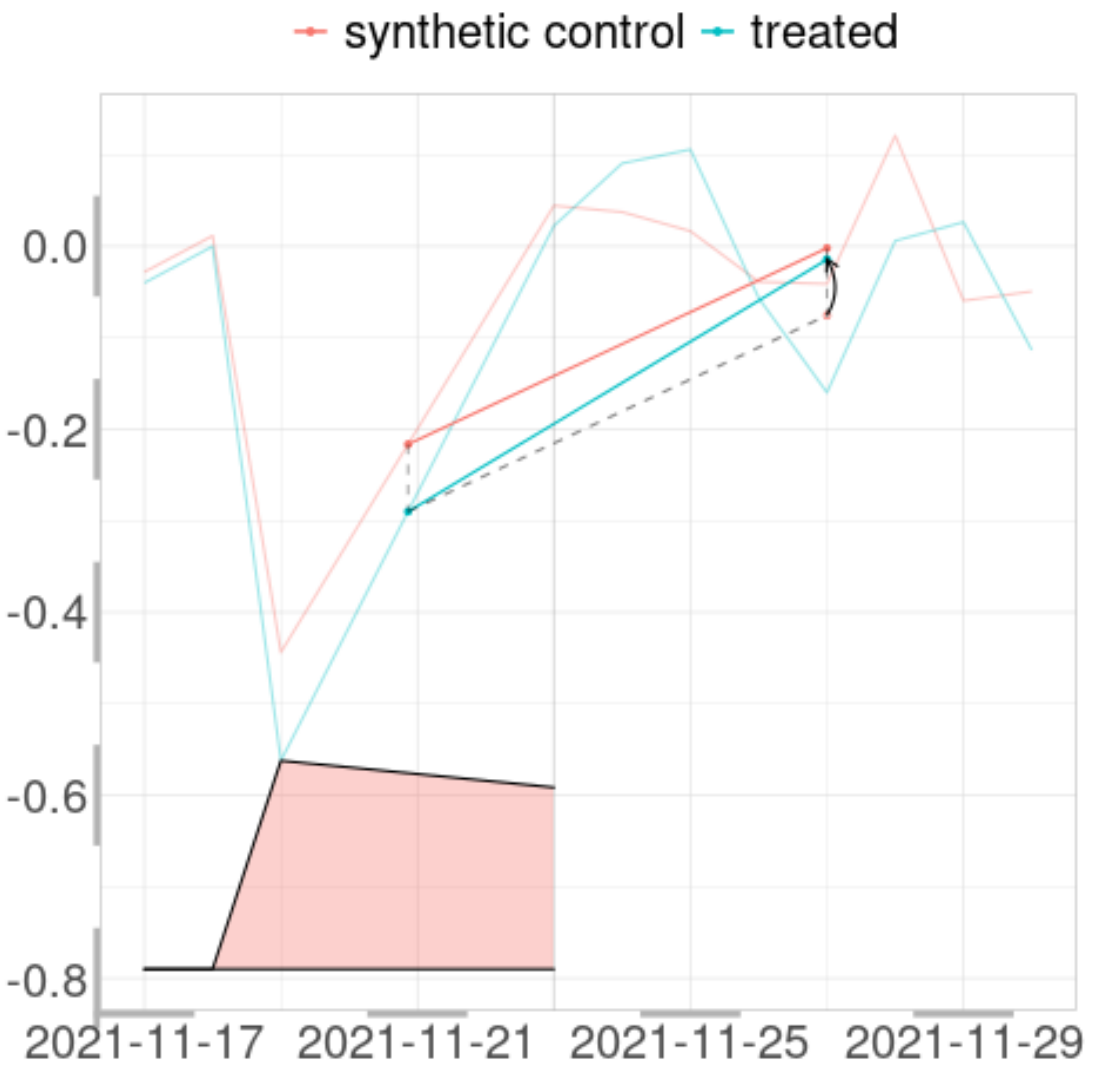
</td></tr>
</table>

**Table S4.** SDID Weights for Analysing Bergen Interventions. The weights assigned to each control region are presented for the applications of SDID to estimate the effect of the two interventions in Bergen on each of the three mobility metrics.

| Date | Control Region | meanDistAway | timeAway | maxDistAway |
|---|---|---|---|---|
| Aug 5 | Oslo | 0.405 | 0.334 | 0.376 |
| | Trondheim | 0.327 | 0.324 | 0.332 |
| | Stavanger | 0.268 | 0.342 | 0.293 |
| Aug 12 | Oslo | 0.397 | 0.321 | 0.376 |
| | Trondheim | 0.303 | 0.349 | 0.288 |
| | Stavanger | 0.300 | 0.330 | 0.336 |

**Table S5.** Visualisation of Mobility Trends in Synthetic Control versus Treated Regions for Analysing Bergen Interventions. The normalised mobility metrics are plotted for a week before and after each intervention for both the synthetic control and treated regions. The synthetic control is a weighted average of the control regions, where the weights are chosen to minimise the difference between the synthetic control and treated region's mobility trends before the intervention.

<table>
<tr><th></th><th>8/5 - Non-Compulsory Measures</th><th>8/12 - Mandate</th></tr>
<tr><td>Mean Dist Away</td><td>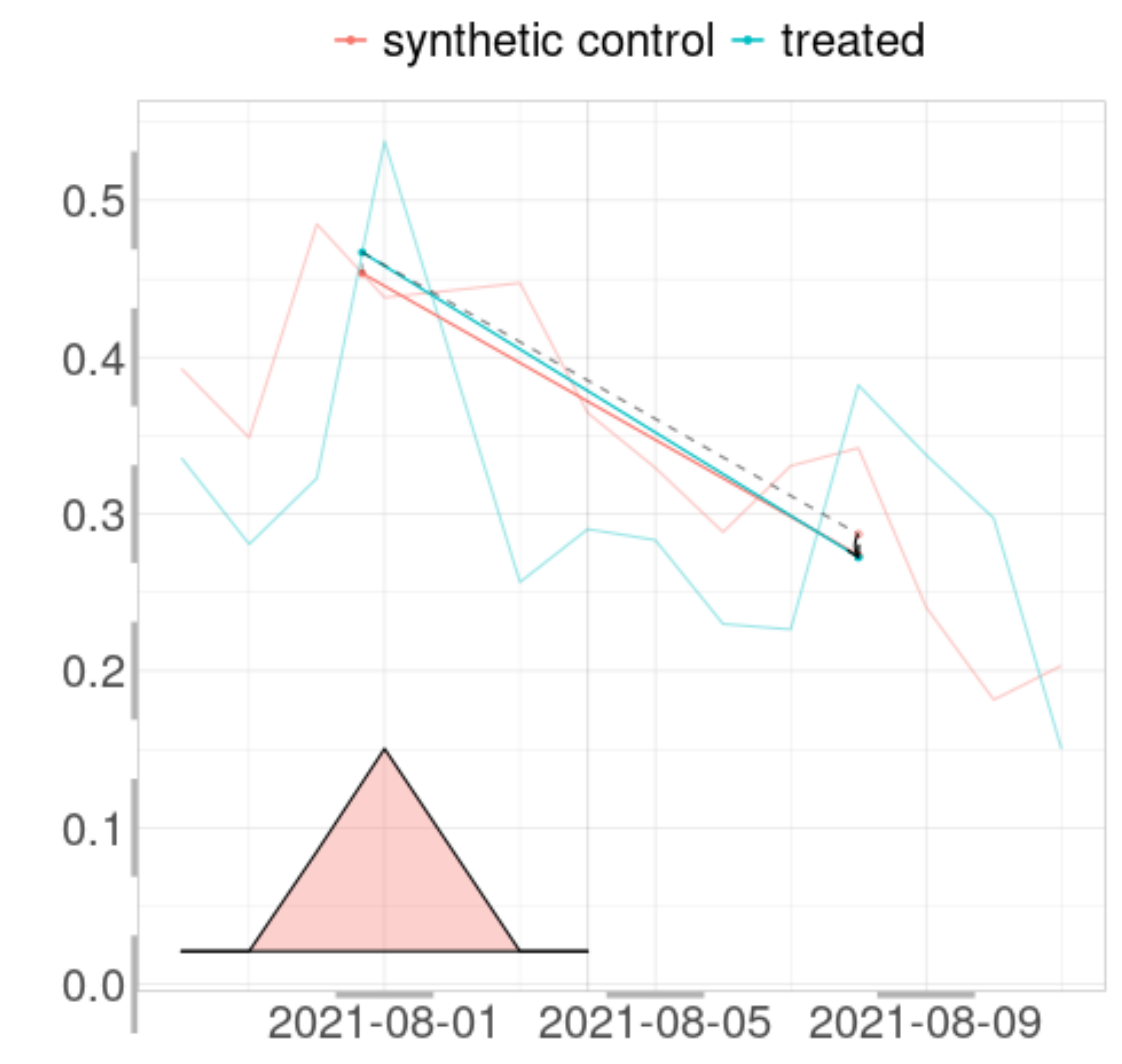
</td><td>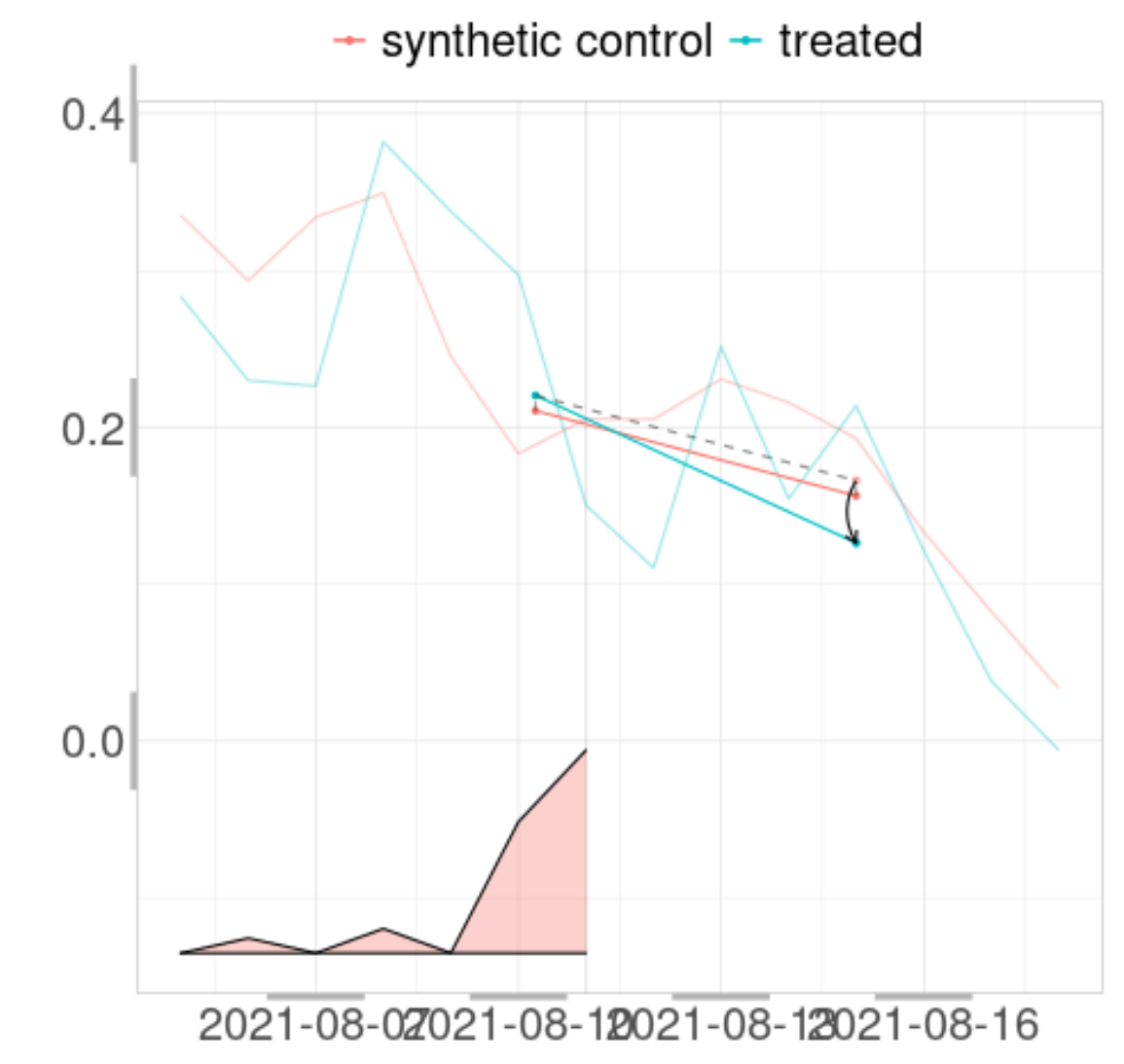
</td></tr>
<tr><td>Time Away</td><td>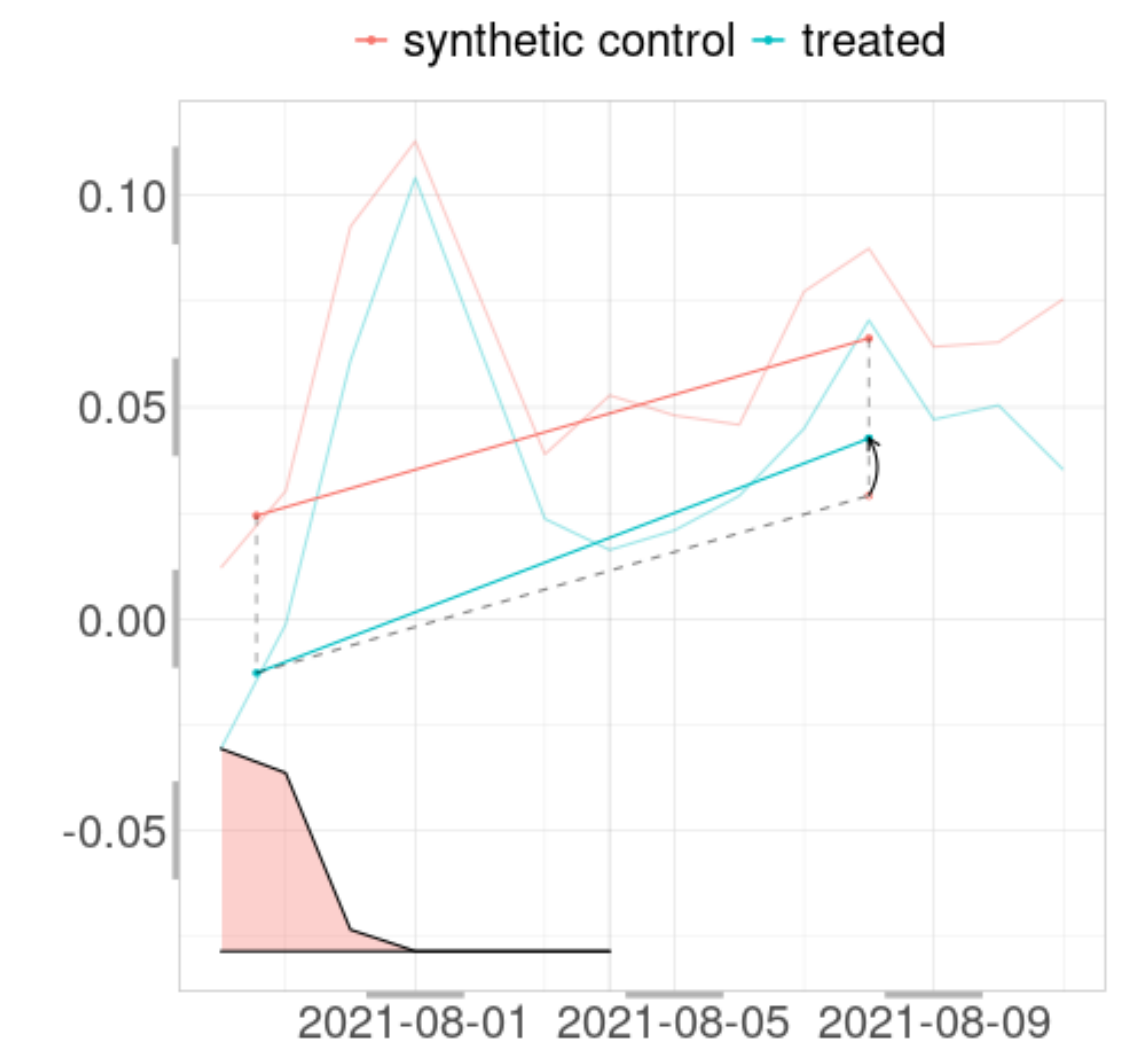
</td><td>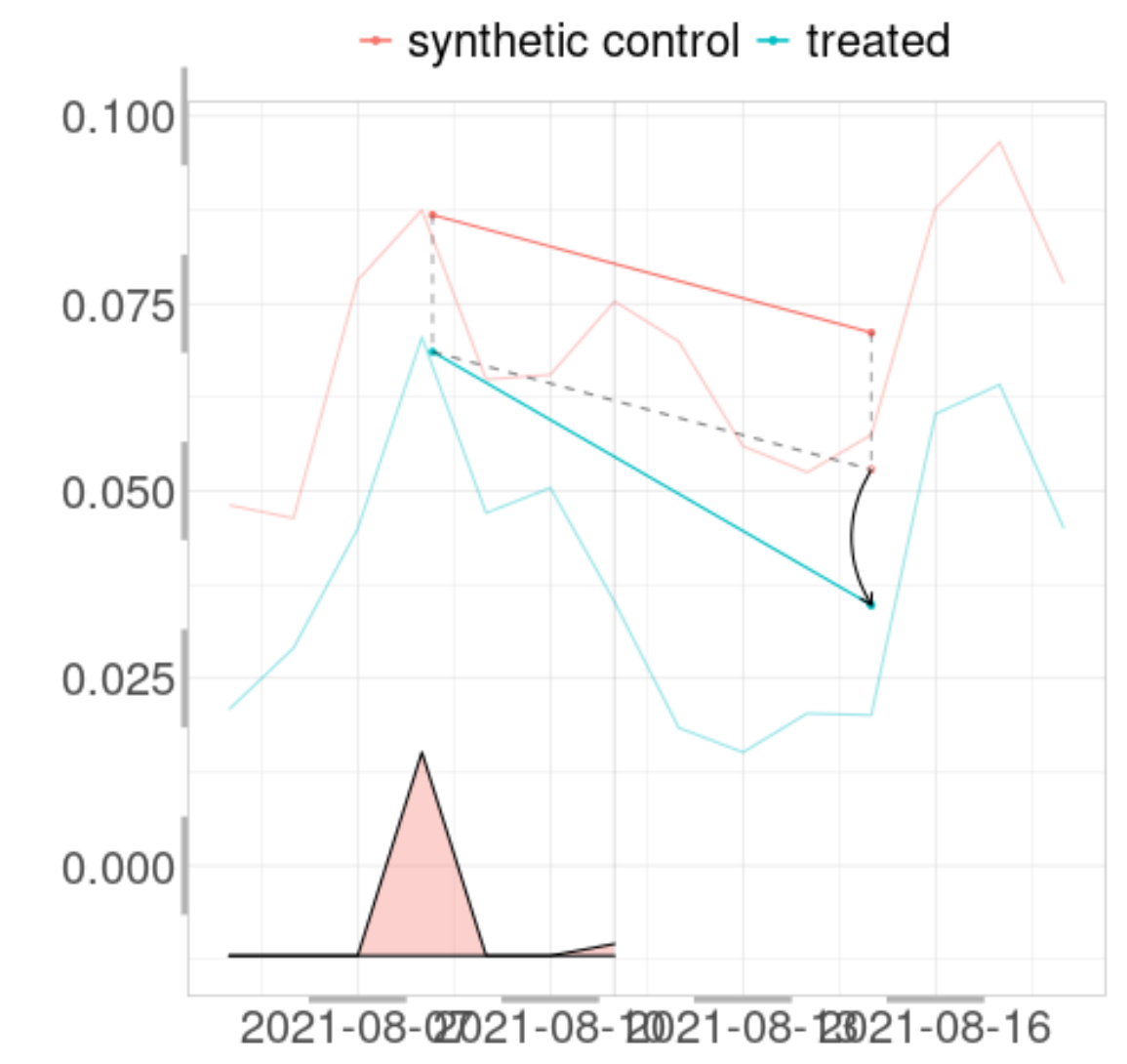
</td></tr>
</table>

| | | |
|---|---|---|
| Max Dist Away | 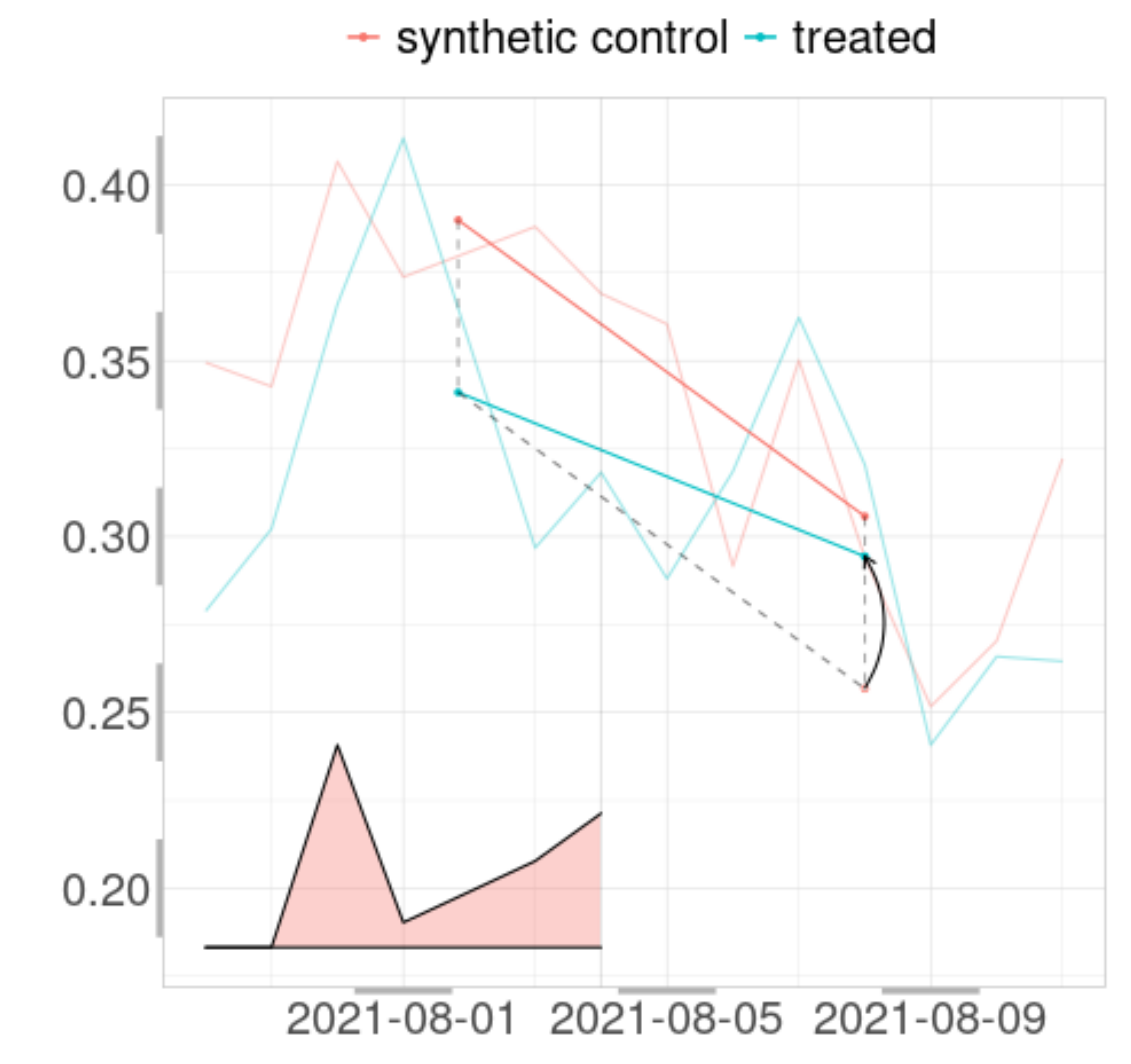 | 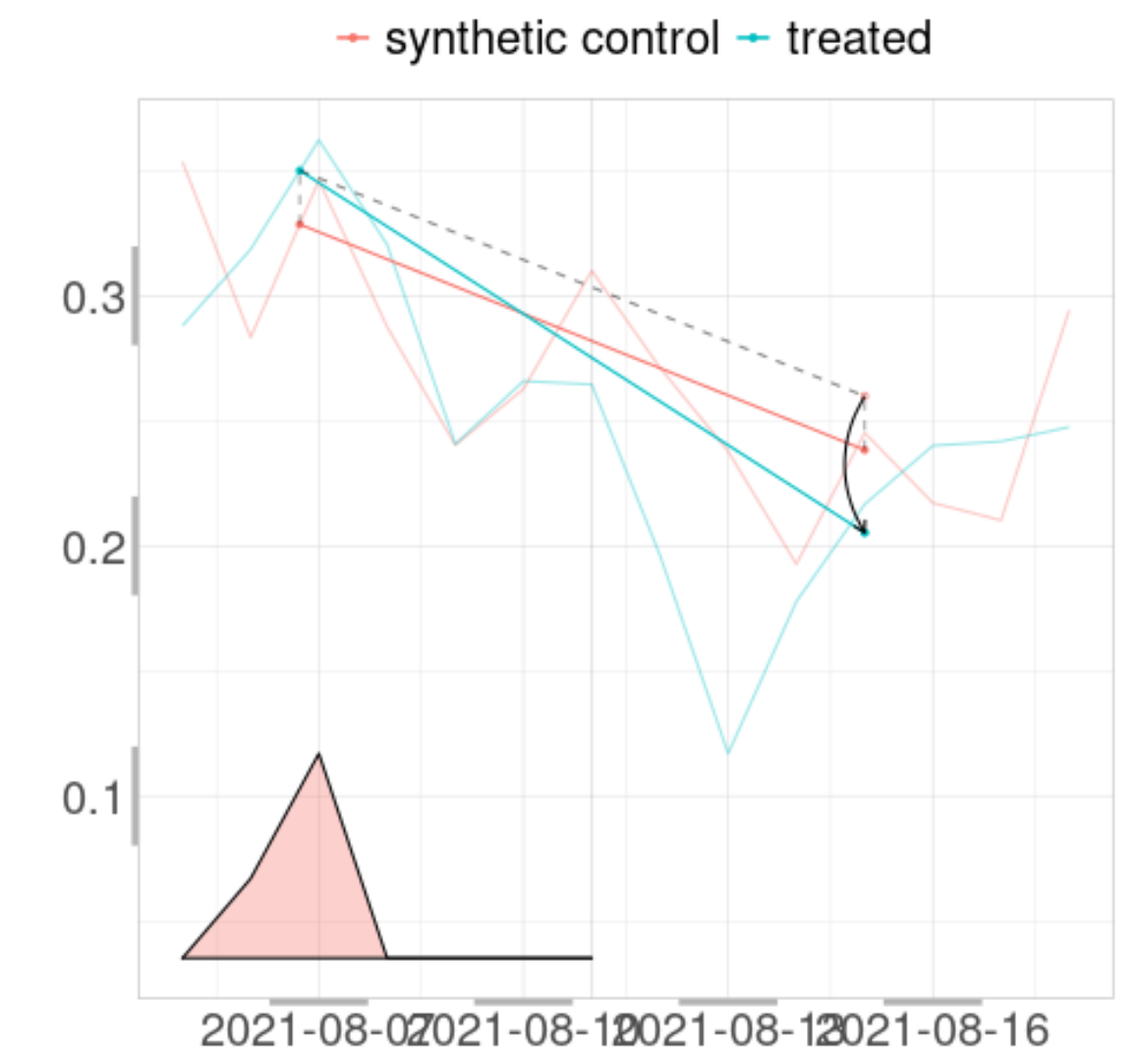 |

**Table S6.** SDID Weights for Analysing Trondheim Interventions. The weights assigned to each control region are presented for the applications of SDID to estimate the effect of the two interventions in Trondheim on each of the three mobility metrics.

| Date | Control Region | meanDistAway | timeAway | maxDistAway |
|---|---|---|---|---|
| Nov 2 | Oslo | 0.494 | 0.301 | 0.347 |
| | Bergen | 0.337 | 0.344 | 0.357 |
| | Stavanger | 0.169 | 0.355 | 0.296 |
| Nov 24 | Oslo | 0.336 | 0.332 | 0.334 |
| | Bergen | 0.330 | 0.334 | 0.332 |
| | Stavanger | 0.334 | 0.334 | 0.334 |

**Table S7.** Visualisation of Mobility Trends in Synthetic Control versus Treated Regions for Analysing Trondheim Interventions. The normalised mobility metrics are plotted for a week before and after each intervention for both the synthetic control and treated regions. The synthetic control is a weighted average of the control regions, where the weights are chosen to minimise the difference between the synthetic control and treated region's mobility trends before the intervention.

| | 11/2 - Non-Compulsory Measures | 11/24 - Mandate |
|---|---|---|
| Mean Dist Away | 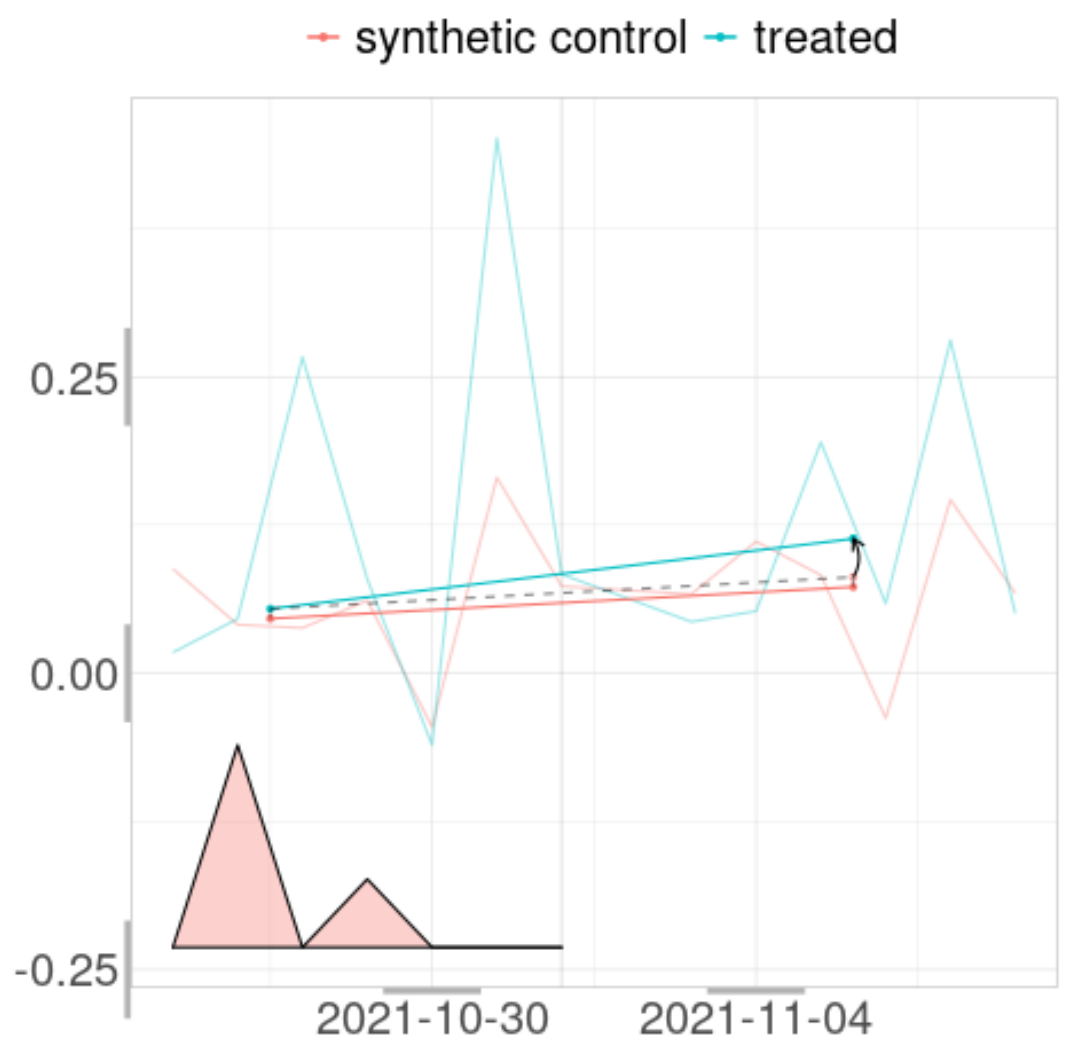  | 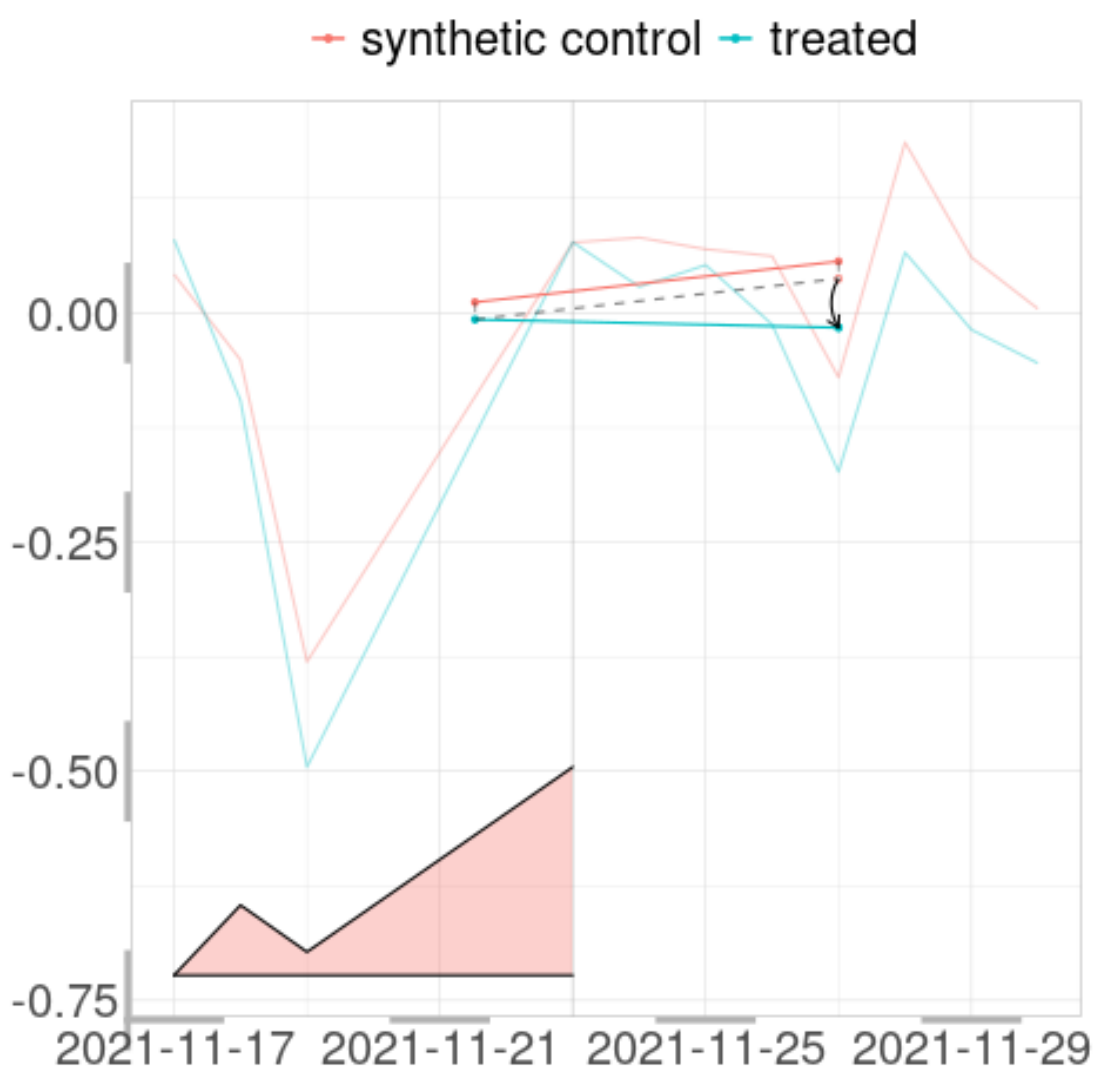  |
| Time Away | 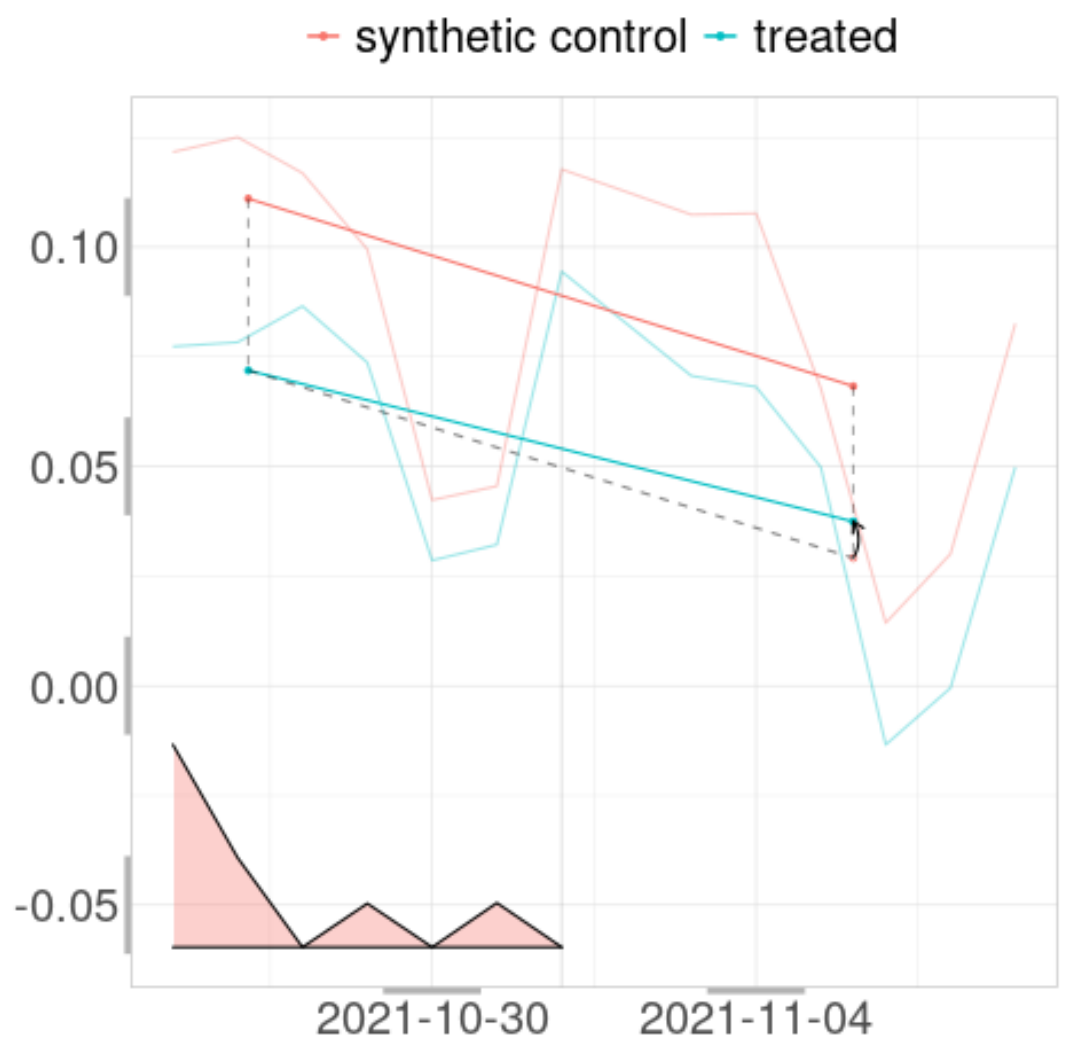  | 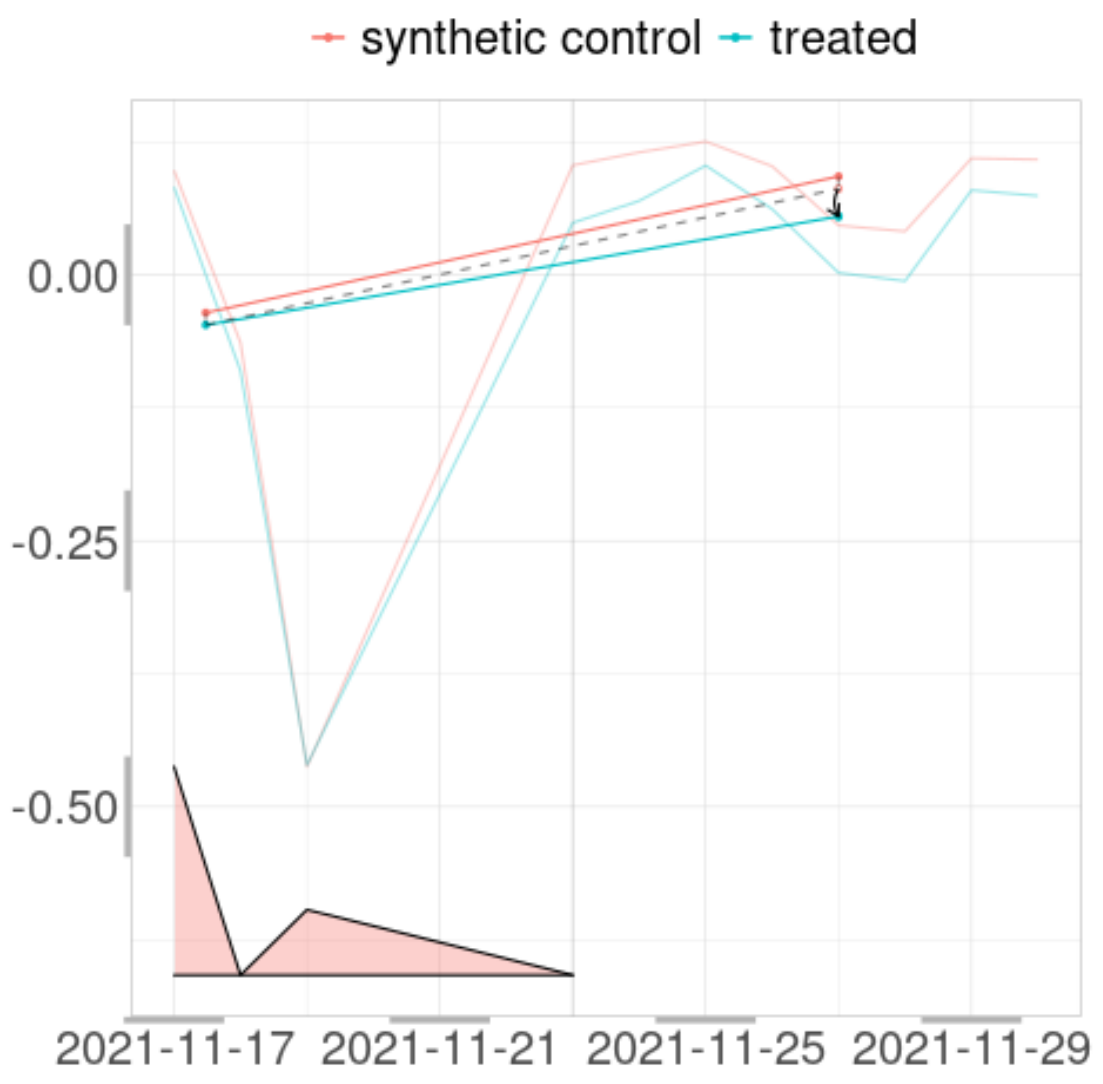  |
| Max Dist Away | 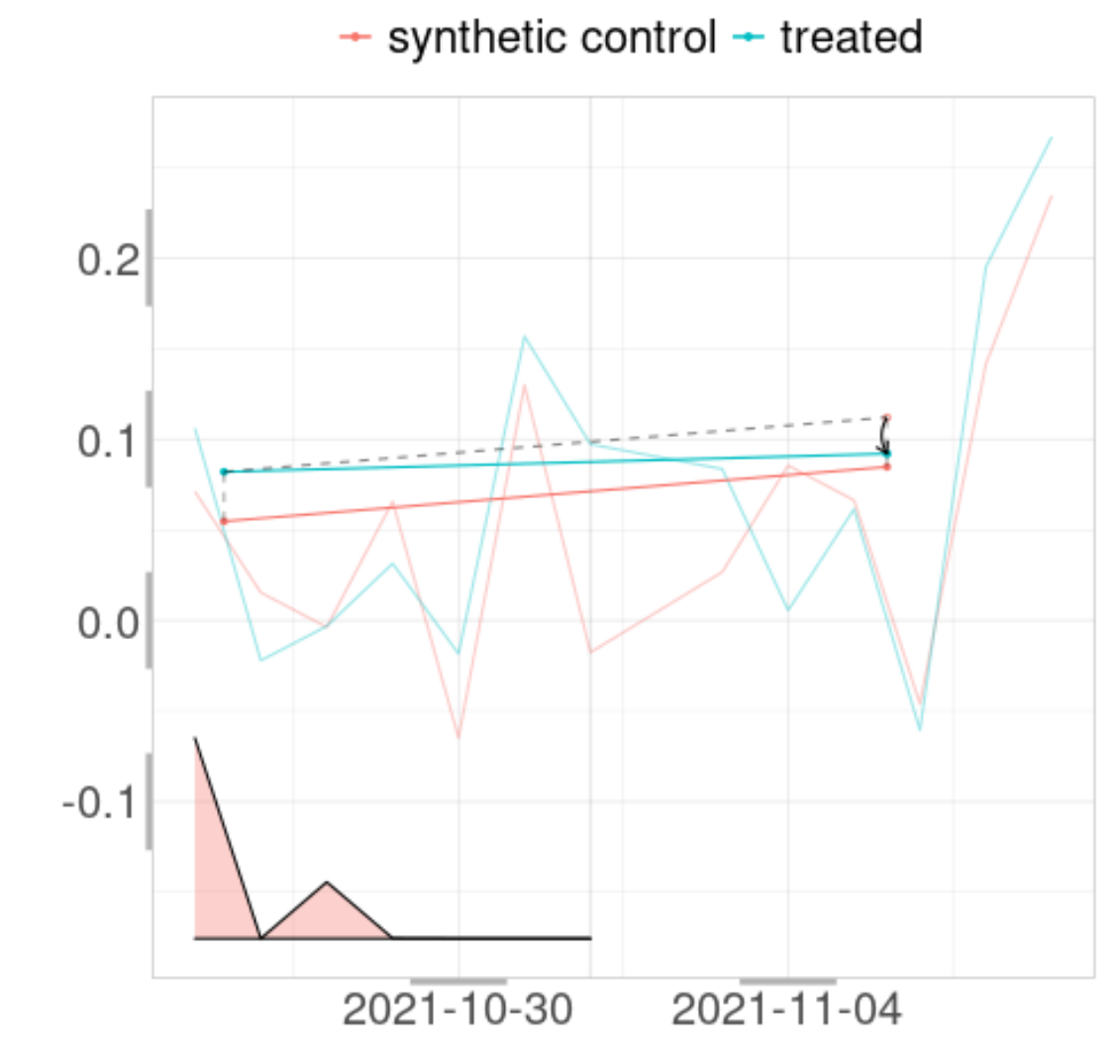 | 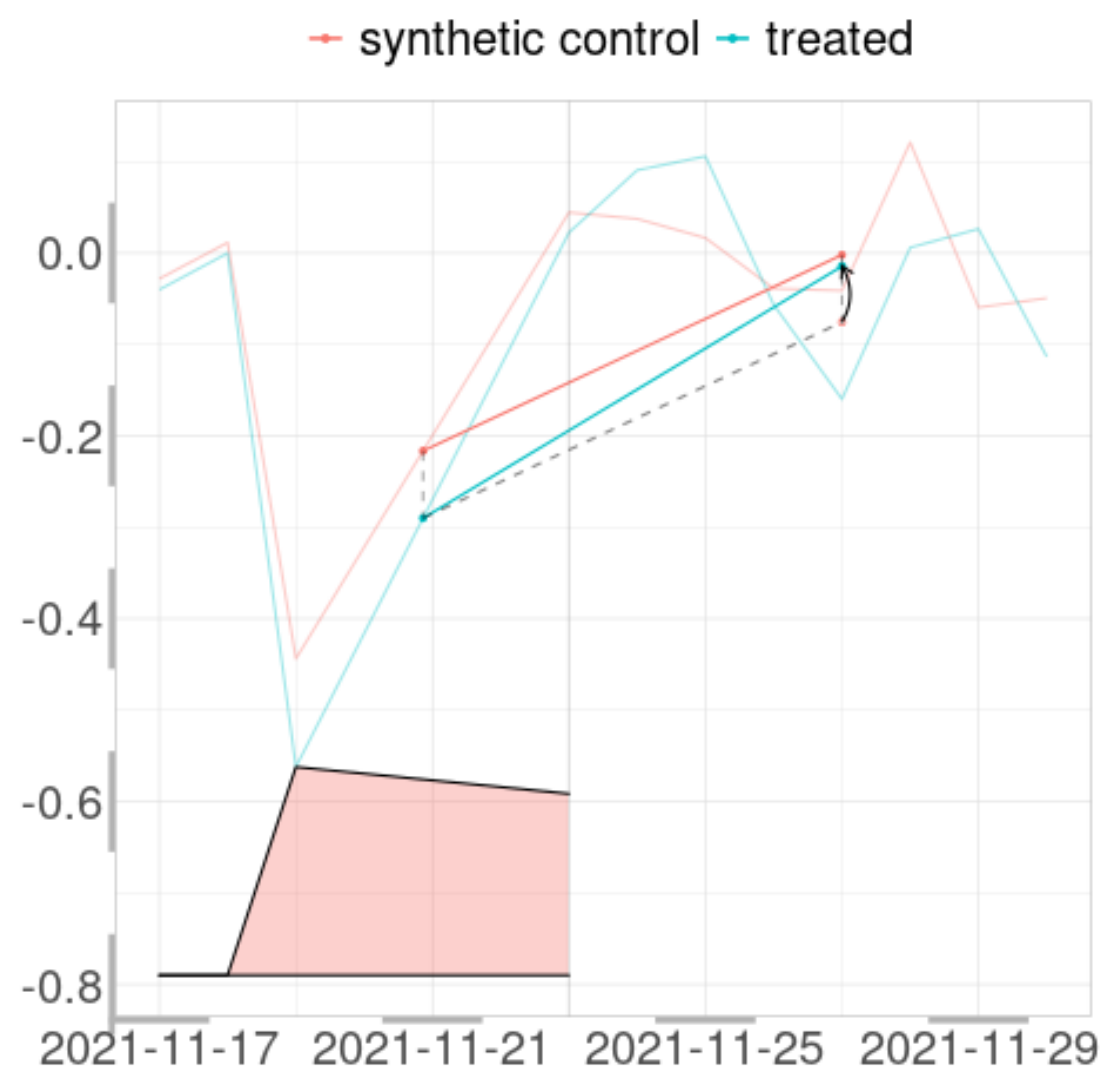 |

## II. Least Populated Municipalities in Each County

The least populated municipalities were chosen based on the number of Telenor users on January 24, 2021. The number of users vary slightly between days.

**Table S8.** List of Municipalities with Smallest Number of Telenor Users in each Fylke in Norway.

| Fylke | Total Number of Telenor Users in Regions Considered | Municipalities |
|---|---|---|
| Rogaland | 36048 | Utsira, Kvitsøy, Bokn, Hjelmeland, Sokndal, Bjerkreim, Lund, Suldal, Sauda, Randaberg, Strand |
| Møre og Romsdal | 57709 | Smøla, Sande, Tingvoll, Aukra, Gjemnes, Fjord, Vanylven, Stranda, Hareid, Aure, Averøy, Sykkylven, Surnadal |
| Nordland | 34974 | Træna, Vevelstad, Røst, Værøy, Moskenes, Vega, Sømna, Flakstad, Bindal, Rødøy, Herøy, Beiarn, Nesna, Hattfjelldal, Grane, Leirfjord, Evenes, Sørfold, Lurøy, Lødingen |
| Viken | 152600 | Aremark, Gjerdrum, Flå, Skiptvet, Marker, Hurdal, Rollag, Hole, Våler, Jevnaker, Hvaler, Nore og Uvdal, Krødsherad, Flesberg, Ål, Enebakk, Råde, Nesbyen, Gol, Rakkestad, Rælingen, Lunner, Hemsedal, Sigdal, Aurskog-Høland |
| Innlandet | 94708 | Folldal, Os, Tolga, Engerdal, Rendalen, Etnedal, Alvdal, Lom, Dovre, Stor-Elvdal, Skjåk, Våler, Vestre-Slidre, Lesja, Vang, Grue, Tynset, Nord-Odal, Vågå, Sør-Aurdal, Eidskog, Øystre-Slidre, Sør-Fron |
| Vestfold og Telemark | 55170 | Siljan, Fyresdal, Tokke, Hjartdal, Kviteseid, Nissedal, Seljord, Drangedal, Nome, Kragerø, Tinn |
| Agder | 45661 | Iveland, Bygland, Åmli, Hægebostad, Gjerstad, Valle, Vegårshei, Evje og Hornnes, Åseral, Birkenes, Sirdal, Bykle |
| Vestland | 62604 | Fedje, Modalen, Solund, Ulvik, Hyllestad, Eidfjord, Aurland, Samnanger, Masfjorden, Lærdal, Austrheim, Vik, Fitjar, Gulen, Fjaler, Askvoll, Etne, Vaksdal, Tysnes, Sveio, Høyanger |
| Trøndelag | 51378 | Leka, Røyrvik, Osen, Namsskogan, Høylandet, Rindal, Tydal, Flatanger, Lierne, Snåase-Snåsa, Holtålen, Frosta, Meråker, Rennebu, Selbu, Overhalla, Grong, Åfjord, Frøya |
| Troms og Finnmark | 30140 | Unjárga - Nesseby, Loppa, Loabák - Lavangen, Berlevåg, Kvænangen, Gratangen, Hasvik, Gamvik, Vardø, Dyrøy, Lebesby, Kárášjohka - Karasjok, Gáivuotna - Kåfjord - Kaivuono, Måsøy, Båtsfjord, Guovdageaidnu - Kautokeino, Salangen, Storfjord - Omasvuotna - Omasvuono, Ibestad |

## F. Multiple Intervention Linear Regression

### I. Model Details

$T_{I,r}$ represents the time points when an NPI was implemented in region $r$, where $r \in R$

$T_{C,r}$ represents the time points that are midpoints of two week time periods when no NPI was implemented in region $r$, where $r \in R$

We denote the set of all time points for region $r$ as $T_r = T_{I,r} \cup T_{C,r}$.

We compute the temperature covariate $P_{r,t}$ using equation 1.

$OriginalTemp_{r,t} =$ temperature in region r on day t

$MeanTemp_r$ = mean temperature in region r from January 10, 2021 to January 9, 2022

$$Temp_{r,t} = \frac{OriginalTemp_{r,t}}{MeanTemp_r},\ P_{r,t} = \frac{\sum_{p=t}^{t+6} Temp_{r,p} - \sum_{p=t-7}^{t-1} Temp_{r,p}}{\sum_{p=t-7}^{t-1} Temp_{r,p}}\ r \in R,\ t \in T_r \quad (1)$$

The national temperature was computed as a weighted average of the temperatures in all counties, weighted by the population of each county.

For each model, we create two interaction covariates between the weekly temperature change and two intervention categories that were significant in an initial regression without the interactions. TempWeeklyChange is interacted with restaurants and shops reopening ("ResShopReopen") and restrictions on guests in private homes ("PHDecr") for the model for meanDistAway, with "ResShopReopen" and face mask recommendations ("FMRec") for the model for timeAway, and with restrictions on alcohol ("LessAlc") and event restrictions ("EventNumDecr") for the model for maxDistAway.

$Int_{i,r,t} = P_{r,t} * X_{i,r,t},\ i \in [A, B]$ where A and B are two manually chosen intervention categories

The models predict the change in mobility after the intervention relative to the initial mobility, computed using equation 2. We utilise daily mean meanDistAway, timeAway, and maxDistAway as the mean metrics.

$$Y_{r,t} = \frac{\sum_{p=t}^{t+6} MeanMetric_{r,p} - \sum_{p=t-7}^{t-1} MeanMetric_{r,p}}{\sum_{p=t-7}^{t-1} MeanMetric_{r,p}}, r \in R,\ t \in T_r \quad (2)$$

For all models, the variance inflation factor was computed for each feature to identify potential multi-collinearities. Almost all variance inflation factor (VIF) values are under 5, which is below the recommended upper threshold [56]. The only exception was a VIF value for face mask recommendations ("FMRec") in the model for timeAway of 6.07, which is still close to the recommended threshold. Therefore, multicollinearity is not a significant issue.

For some intervention time points, the time period of comparison overlapped with either another intervention or a major holiday period, defined as the time around Easter, Christmas, and New Year's. We manually created new periods of analysis for this intervention and used equation 3 to compute the weekly mobility change for these interventions. Table 10 lists the time points for which equation 3 was used to compute the outcome.

$$Y_{r,t} = \frac{\sum_{p=afterStart}^{afterEnd} MeanMetric_{r,p} - \sum_{p=beforeStart}^{beforeEnd} MeanMetric_{r,p}}{\sum_{p=beforeStart}^{beforeEnd} MeanMetric_{r,p}}, r \in R,\ t \in T_r \quad (3)$$

**Table S9.** Time Points with Modified Comparison Time Frames. For some interventions, time frames for the periods before and after the intervention of interest were modified to avoid overlap with holiday periods or other interventions. Equation 3 is used to compute the change in mobility for these intervention time points.

| Time point | beforeStart | beforeEnd | afterStart | afterEnd |
|---|---|---|---|---|
| 3-25 | 3-18 | 3-24 | 4-8 | 4-14 |
| 3-27 | 3-20 | 3-26 | 3-27 | 3-30 |
| 3-31 | 3-27 | 3-30 | 3-31 | 4-6 |
| 12-3 | 11-26 | 12-2 | 12-3 | 12-8 |
| 12-9 | 12-3 | 12-8 | 12-9 | 12-14 |
| 12-15 | 12-9 | 12-14 | 1-2 | 1-8 |

## II. Intervention Categories to Intervention Dates

**Table S10.** Mapping of Intervention Categories to Dates of Implementation for each Region. Interventions are included both nationally and from Oslo, Bergen, Trondheim, Tromsø, and Stavanger. We do not include local December interventions around the 3rd, 9th, or 15th, as there are so many national interventions that were added at that time point.

| Feature in Model | Corresponding Interventions | National | Oslo | Bergen | Trondheim | Tromsø | Stavanger |
|---|---|---|---|---|---|---|---|
| StricterWH | Work from home required, Normal office to Work from home recommended | 12/15, 12/3, 12/9 | | 2/7 | 2/10, 6/1 | 11/9, 10/28 | 4/16 |
| WHEased | Work from home removed or becomes less mandatory | 6/20 | 7/5 | 2/21, 5/31 | 6/22 | | 5/6 |
| FMReq | Face mask required | 12/9, 12/15 | | 2/7, 4/19, 8/12 | 11/24, 5/19, 8/26 | 11/9, 3/5 | 9/1 |
| FMRec | Face mask recommended | 3/25, 12/3 | 11/24 | 8/5 | 11/2 | 7/2, 10/28 | |
| FMEased | Face mask measures eased | | | 9/1 | 6/22, 9/21 | 3/19, 7/13, 11/30 | 9/15 |
| StricterMeterRule | One metre rule added and recommendation to reduce close contacts | 3/25, 12/3 | | | | 10/28 | |
| LessSocDist | Removing recommendation to social distance or easing metre rule | 9/25 | 6/16 | | | | |
| PHDecr | Private home limit decreased or made more mandatory | | 3/2 | 2/7, 4/19, 8/5, 8/12 | 6/1 | 3/5 | |
| PHIncr | Private home limit increased or made less mandatory | 2/3, 4/16, 6/20 | 6/16 | 5/12, 2/21, 5/31, 9/1 | 6/22 | 3/19 | |
| LessAlc | Adding pouring stops or requirement of 1 metre at bars | 3/25, 12/9, 12/15 | | 2/7, 5/12 | 6/1, 2/10 | | |
| AddAlc | Easing pouring stops or 1 metre at pubs | 4/16, 5/27, 6/20, 9/25 | 5/26, 6/16, 7/5 | 2/21, 5/31 | 2/2, 6/22 | | 5/6 |
| EventNumDecr | Number of people at events decreased, or event restrictions | 3/25, 12/9, 12/15 | 3/2 | 2/7, 4/19 | 9/2 | 11/9 | 4/16 |
| EventNumIncr | Number of people at events increased, or less event restrictions | 2/3, 2/23, 4/16, 5/27, 6/20, 9/4, 9/25 | 5/26, 6/16 | 5/31, 2/21 | | 11/30 | 5/6, 7/8 |
| SchoolLimitsEased | less restrictions on schools | 4/16 | 2/3, 4/19 | 2/21 | | | |
| StricterSchoolLimits | more restrictions on schools | | 3/2 | 2/7, 4/19 | | | |
| GymsClosed | gyms closed | | | 3/27 | | 3/5 | |
| GymsReopen | gyms reopen | | | 3/31 | | 3/19 | |
| ResShopReopen | Restaurants and businesses can reopen | | 2/3, 5/6, 5/26 | | | | |
| ResShopClose | Restaurants and shops closed | | 3/2 | | 2/10 | | |

### III. Creation of Control Time Points

The algorithm to identify non-intervention time points involves the following steps.

1. Start with a set of all dates, D. For each date t in the list of all national and local intervention dates, remove all dates from t-7 to t+6, inclusive. Only use national interventions when the region of interest is Norway.
2. Remove dates related to holidays. I removed one week before and after 12-28 and 4-5 to account for the winter and Easter holidays.
3. Iterate through dates in D in chronological order. For each date t, add t to the control time point list ($T_{C,r}$) if the t-7 and t+6 are still in D. Then, remove all dates from t-7 to t-6, inclusive from D.